\documentclass [superscriptaddress,amsmath,prl,twocolumn,amssymb]{revtex4-1}
\usepackage{graphicx}
\usepackage{dcolumn}
\usepackage[none]{hyphenat} 
\usepackage{braket}
\usepackage{multirow}
\usepackage{booktabs}
\usepackage{bm,color,upgreek}
\usepackage{braket}

\begin{document}

\title{The Berry curvature dipole in Weyl semimetal materials: an \textit{ab~initio} study}

\author{Yang Zhang}
\affiliation{Max Planck Institute for Chemical Physics of Solids, 01187 Dresden, Germany}
\affiliation{Leibniz Institute for Solid State and Materials Research, IFW Dresden, 01069 Dresden, Germany}
\author{Yan Sun}
\affiliation{Max Planck Institute for Chemical Physics of Solids, 01187 Dresden, Germany}
\author{Binghai Yan}
\email{binghai.yan@weizmann.ac.il}
\affiliation{Department of Condensed Matter Physics, Weizmann Institute of Science, Rehovot, 7610001, Israel}

\begin{abstract}
Noncentrosymmetric metals are anticipated to exhibit a $dc$ photocurrent in the nonlinear optical response caused by the Berry curvature dipole in momentum space.
Weyl semimetals (WSMs) are expected to be excellent candidates for observing these nonlinear effects because they carry a large Berry curvature concentrated in small regions, i.e., near the Weyl points.
We have implemented the semiclassical Berry curvature dipole formalism into an \textit{ab~initio} scheme and investigated the second-order nonlinear response for two representative groups of materials: the TaAs-family type-I WSMs and MoTe$_2$-family type-II WSMs.
Both types of WSMs exhibited a Berry curvature dipole, in which type-II Weyl points are usually superior to the type-I because of the strong tilt.
Corresponding nonlinear susceptibilities in several materials promise a nonlinear Hall effect in the $dc$ field limit, which is within the experimentally detectable range. 
\end{abstract}

\maketitle

\textit{Introduction --}
The Weyl semimetal (WSM)~\cite{Wan2011,volovik2003universe,Burkov2011,Hosur2013,Yan2017,Armitage2017}
 is a topological state characterized by linear band crossing points called Weyl points near the Fermi energy.
WSM materials such as the TaAs-family pnictides~\cite{Weng2015,Huang2015} and MoTe$_2$~\cite{Soluyanov2015WTe2,Sun2015MoTe2} 
 have recently been discovered primarily by observation of the unique Fermi arcs of surface states through angle-resolved photoemission spectroscopy~\cite{Lv2015TaAs,Xu2015TaAs,Yang2015TaAs,Deng2016,Jiang2016,Huang2016}.
Because Weyl points are monopole sources or drains of the Berry curvature of Bloch wave functions in momentum space, a WSM can exhibit an anomalous Hall effect when breaking the time-reversal symmetry (TRS)~\cite{Xu2011,Yang2011QHE,Burkov2014} or a spin Hall effect~\cite{Sun2016}, as a linear response to an external electric field.
Recent theoretical~\cite{Hosur2015,Sodemann2015,Morimoto2016a,Taguchi2016,Morimoto2016,Chan2016,Ishizuka2016,deJuan2017,Chan2017,Rostami2017} and experimental~\cite{Wu2017nonlinear,Ma2017,Sun2016CPGE,Chi2017} studies have revealed giant nonlinear optical responses in 
inversion-symmetry-breaking WSMs, such as the photocurrent from the circular photogalvanic effect (CPGE), second harmonic generation (SHG), and nonlinear Hall effect.
These nonlinear effects can be much stronger in WSMs than traditional electro-optic materials owing to the large Berry curvature~\cite{Moore2010,Deyo2009,Sodemann2015}.

Very recently, the semiclassical approach has been used to describe the intraband contributions to CPGE and SHG as a Berry phase effect~\cite{Moore2010,Deyo2009} by a geometric quantity: the Berry curvature dipole (BCD)~\cite{Sodemann2015}. At the $dc$ limit, the photocurrent remains finite as a transverse Hall-like current, i.e., a nonlinear Hall effect~\cite{Sodemann2015}. These nonlinear effects originate from the intraband resonant transitions at a low frequency in a noncentrosymmetric metal.
Although they have played an important role in predicting topological materials and estimating their linear-response properties, there is still a lack of \textit{ab~initio} studies on the nonlinear optical effects of WSMs to quantitatively reveal the role of the Weyl points in realistic materials~\cite{Wu2017nonlinear}.
The nonlinear response is usually computed with mixed interband and intraband transitions for conventional semiconductors~\cite{Sipe2000,Young2012}, but an \textit{ab~initio} scheme with the Berry phase formalism is required to understand WSMs.

We studied the BCD of WSM materials, i.e., TaAs-family type-I and MoTe$_2$-family type-II WSMs, and estimated their nonlinear optical responses by \textit{ab~initio} calculations combined with the semiclassical approach.
Both types of WSMs exhibit a large BCD near the Weyl point that is orders of magnitude larger than that of conventional materials.
As a Fermi surface property, the BCD favors tilted Weyl cones. Thus, the type-II WSM is usually superior to the type-I WSM.
Further, we found that some small gap regions with highly concentrated Berry curvature can also contribute to a large dipole in the absence of Weyl points.
We made an order-of-magnitude estimate of the nonlinear Hall effect for titled materials, 
which is experimentally accessible.

\textit{Semiclassical theory --} We first overview previous theoretical work on the nonlinear optical response described by the Berry curvature~\cite{Moore2010,Deyo2009,Sodemann2015,Morimoto2016}.
For the CPGE, the oscillating electric field $E_c(t)=Re\{ \mathcal{E}_c e^{i \omega t}\}$ of circularly polarized light induces a $dc$ photocurrent $\textbf{\textit{j}}^{(0)}$ as a second-order nonlinear optical effect: $\textit{j}_a^{(0)}=\chi_{abc} \mathcal{E}_b \mathcal{E}_c^*$. Similarly, the SHG is described by the second-harmonic current response $\textbf{\textit{j}}^{(2\omega)} e^{2 i\omega t}$ to a linearly polarized light, where $\textit{j}_a^{(2\omega)}=\chi_{abc} \mathcal{E}_b \mathcal{E}_c$.
At the $dc$ limit of a linearly polarized field, the nonlinear Hall effect is characterized by a transverse current:
$\textit{j}_a= 2 \textit{j}_a^{(0)} |_{\omega \rightarrow 0} = 2 \chi_{abb} |\mathcal{E}_b|^2$.
For a material with TRS, the nonlinear response tensor $\sigma$ has been theoretically obtained as a Berry phase effect~\cite{Moore2010,Deyo2009} and very recently further described by the BCD~\cite{Sodemann2015} as follows:
\begin{eqnarray}
\label{eq:chi}
\chi_{abc} &=& - \varepsilon_{adc} \frac{e^3\uptau}{2 \hbar^2 (1+ i \omega \uptau)} D_{bd}\\
\label{eq:dipole}
D_{bd} &=& \int_k f_0 \frac{\partial{\Omega_d}}{\partial{k_b}},
\end{eqnarray}
where $D_{bd}$ is the BCD, $\Omega_d$ is the Berry curvature, $f_0$ is the equilibrium Fermi--Dirac distribution, $\uptau$ refers to the relaxation time approximation in the Boltzmann equation, $\varepsilon_{adc}$ stands for the third rank Levi--Civita symbol, and $\hbar$ is the reduced Planck constant. $D_{bd}$ is a Fermi surface effect that is intrinsic to the band structure and becomes dimensionless in three dimensions. We define the BCD density in the $k$-space as $d_{bd} \equiv f_0 \frac{\partial{\Omega_d}}{\partial{k_b}}$. Because $d_{bd}$ is odd to the space inversion, $D_{bd}$ vanishes when inversion symmetry appears.
While they were obtained with the semiclassical theory, Eqs.~\ref{eq:chi} and ~\ref{eq:dipole} can also be derived by a fully quantum theoretical treatment with the Floquet formalism~\cite{Morimoto2016}.

\textit{Ab initio calculation methods --} We performed \textit{ab~initio} density-functional theory (DFT) calculations for the bulk materials and projected Bloch wave functions to atomic-like local Wannier functions with the full-potential local-orbital (\textsc{fplo}) program~\cite{koepernik1999} within the generalized gradient approximation (GGA)~\cite{perdew1996}.
For a material, we obtained the tight-binding Hamiltonian $\hat{H}$. Note that $\hat{H}$ inherits exactly all symmetries of the system, which is crucial for accurate evaluation of the BCD from the Berry curvature $\mathbf{\Omega}$ in a differential manner~\cite{note}. The Berry curvature~\cite{Xiao2010} of the $n$th band can be calculated according to $\hat{H}$:
\begin{equation}
\label{eq:berry}
\begin{aligned}
\Omega^n_a(\mathbf{k})= \varepsilon_{abc} 2i \sum_{m \ne n} \dfrac{<n|\partial_{k_b}\hat{H}|m><m|\partial_{k_c}\hat{H}|n>}{(\epsilon_{n}-\epsilon_{m})^2},
\end{aligned}
\end{equation}
where $\epsilon_{n}$ and $|n>$ are eigenvalues and eigen wave functions, respectively, of $\hat{H}$ at the momentum $\mathbf{k}$. $\Omega^n_a$ runs over occupied bands in Eq.~\ref{eq:dipole}, where $\Omega_d = \sum_{n} \Omega^n_d$.

\begin{figure}[htbp]
\begin{center}
\includegraphics[width=0.5\textwidth]{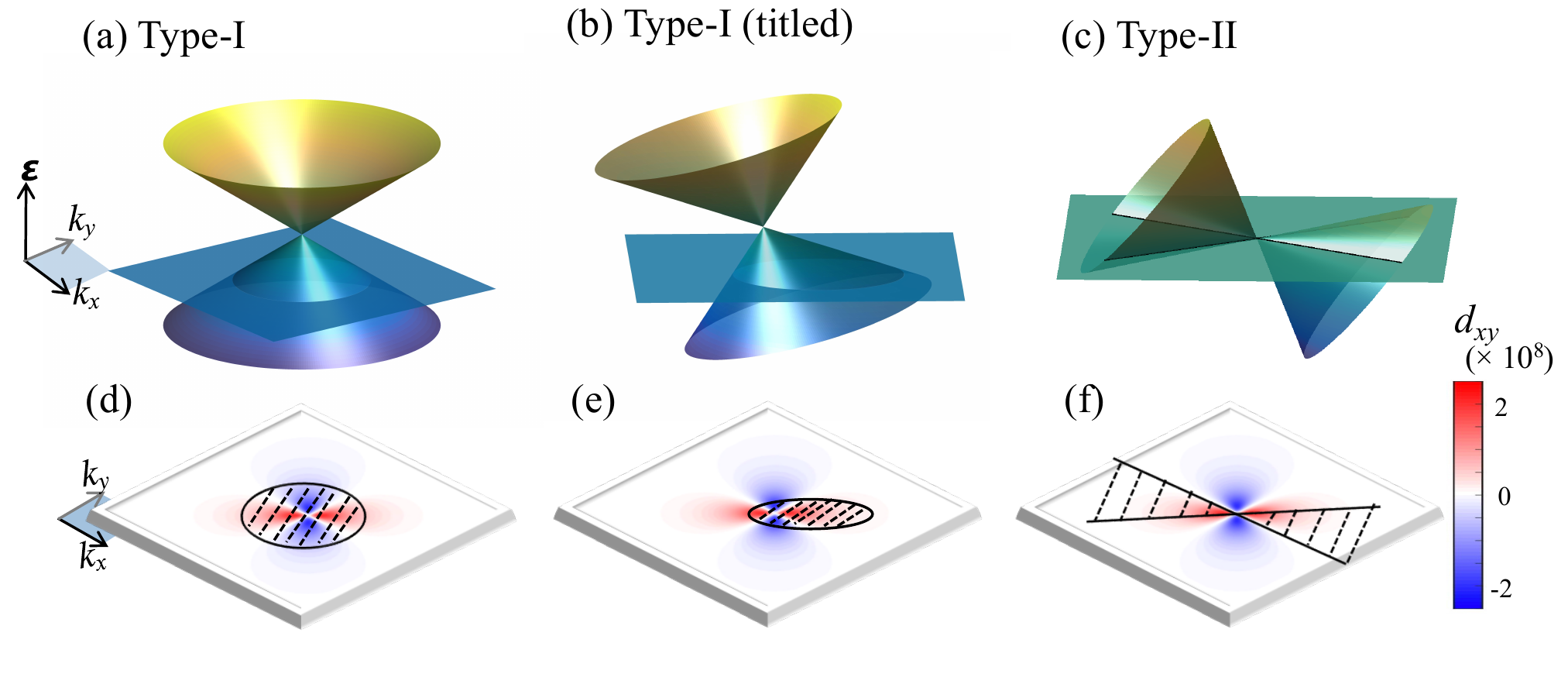}
\end{center}
\caption{
The Weyl cones and the dipole moment distribution of the Berry curvature.
(a) A standard type-I Weyl cone without tilting. The Fermi energy is indicated by the horizontal plane and the Fermi surface is a circle.
Corresponding dipole moment of the Berry curvature is shown in (d) near the Weyl point.
(b) A type-I Weyl cone with a slight tilting and corresponding dipole moment in (e).
(c) A type-II Weyl cone with a strong tilting and corresponding dipole moment in (f).
Near the Weyl point, the dipole moment exhibits a symmetric $k_x k_y$-type distribution when the Fermi energy crosses a type-I Weyl point
and thus, it is summed to be zero as integrating over the $k$-space.
In (d) and (e), the circle with a shadow region indicates the unoccupied bands that do not contribute to the integral of the dipole moment.
The blue and red colors show negative and positive values of the dipole moment.
In (f), the shadowed regions stand for the unoccupied hole pocket and the occupied electron band,
both of which are deducted from the integral of the dipole moment.
}
\label{model}
\end{figure}

\textit{Simple effective model of Weyl points -- }
Before visiting specific WSM materials, we investigated the BCD for a simple Weyl Hamiltonian to reach a qualitative but inspiring understanding:
\begin{equation}
\label{eq:Weyl}
H_{Weyl}(\mathbf{q})=\hbar v_t q_t \sigma_0+ \hbar v_F \mathbf{q} \cdot \mathbf{\sigma},
\end{equation}
where $\mathbf{q}$ is the momentum with respect to the Weyl point, $\mathbf{\sigma}$ is the Pauli matrix vector, $v_F$ is the Fermi velocity of an isotropic Weyl cone without tilt, $v_t$ represents the tilting velocity, and $q_t$ is the tilting momentum along the $\hat{t}$ direction.
The tilt of the Weyl cone is characterized by the ratio $| v_t / v_F |$, where $| v_t / v_F |<1$ stands for a type-I Weyl cone and $| v_t / v_F |>1$ stands for a type-II one.
Because the Berry curvature is $ \Omega (\mathbf{q})=\frac{\mathbf{q}}{2 q^3} $ for the lower cone, we can analytically obtain $d_{xy}$, for example, without loss of generality:
\begin{equation}
\label{eq:dxy}
d_{xy}= \frac{\partial \Omega_y}{\partial q_x} = \frac{3 q_x q_y}{2 q^5}.
\end{equation}
We note that $ \Omega$ and $d_{xy}$ are independent of the tilt and reverse sign for the upper cone. However, the shape of the Fermi surface is sensitive to the tilt.

The $d_{xy}$ exhibits $xy$-type symmetry near the Weyl point (Eq.~\ref{eq:dxy}), which resembles a ``$d_{xy}$-type'' atomic wave function in real space.
For a type-I WSM, $D_{xy}$ diminishes when $E_F$ crosses the Weyl point because the integral of $d_{xy}$ leads to zero owing to the $xy$-type symmetry. This is fully consistent with the fact that $D_{xy}$ vanishes as the Fermi surface shrinks to a point at the Weyl point.
When $E_F$ lies either below or above the Weyl point, the Fermi surface region is effectively subtracted from the sum over the lower cone.
If the type-I Weyl cone has no tilt (see Fig. 1a), the Fermi surface is centered to the Weyl point.
Thus, $d_{xy}$ outside the Fermi surface region is still highly symmetric and summed up to be zero.
If the type-I Weyl cone has a tilt along some arbitrary direction (see Fig. 1b), the Fermi surface center is shifted away from the Weyl point.
Consequently, $d_{xy}$ outside the Fermi surface region becomes asymmetric, which leads to a nonzero net $D_{xy}$.
For a type-II Weyl cone (see Fig. 1c), the Fermi surface naturally breaks the $xy$-type symmetries of $d_{xy}$ and thus presents a nonzero $D_{xy}$. We can simply summarize these optimal conditions for a large $D_{xy}$ near a single Weyl point: (i) For a type-I Weyl point, a tilt is necessary, which is common for WSM materials. Because $d_{xy}$ is highly concentrated near the Weyl point, $E_F$ should stay close enough to the Weyl point.
(ii) The type-II Weyl point may exhibit large $D_{xy}$, even when $E_F$ crosses it. Although the large tilt of Weyl points was also predicted to generate photocurrents by Chan et al.~\cite{Chan2017}, they referred to the resonant transition between occupied and empty bands of the Weyl cone, which is different from the current finding in the low-frequency intraband transition.

Further, we point out that a pair of Weyl points that are the $\mathcal{M}_x$, $\mathcal{M}_y$  or TRS partners contribute the same $D_{xy}$ because $d_{xy}$ is even to $\mathcal{M}_x$, $\mathcal{M}_y$  or TRS. Therefore, multiple Weyl points related to TRS and mirror symmetries multiply their contributions to the BCD instead of compensating for each other.

\begin{figure}[htbp]
\begin{center}
\includegraphics[width=0.5\textwidth]{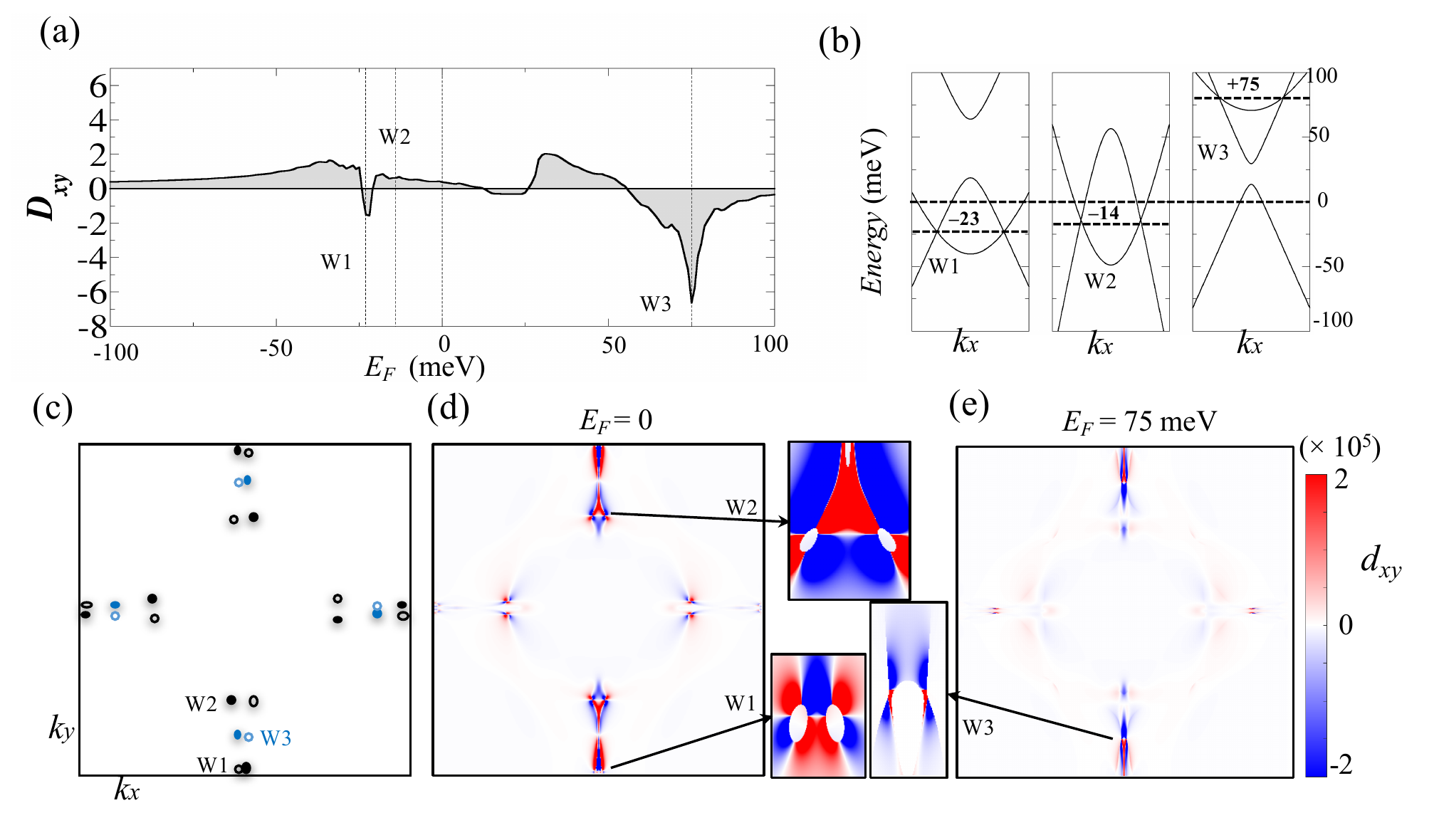}
\end{center}
\caption{Calculated Berry curvature dipole $D_{xy}$ for TaAs.
(a) The Fermi energy ($E_F$) dependence of $D_{xy}$. $E_F=0$ corresponds to the charge neutral point.
(b) The band dispersions crossing a pair of Weyl points. Three types of Weyl points, W1, W2 and W3, are shown.
(c) The projection of three types of Weyl points to the $k_x k_y$ plane by integrating $d_{xy}$ over $k_z$.
The distribution of $d_{xy}$ at (d) $E_F = 0$ and (e) $E_F = 75 $ meV crossing W3.
Red and blue represent positive and negative values of $d_{xy}$, respectively, in the arbitrary unit.
As insets, $d_{xy}$ distributions near some Weyl points are shown in the $k_x k_y$ plane without integrating the $k_z$ direction.
}
\label{structure}
\end{figure}

\textit{Realistic materials --}
We investigated two representative family of materials with inversion symmetry breaking: (Ta, Nb)(As, P) as type-I WSMs and (Mo, W)Te$_2$ as type-II WSMs.
For a given material, the BCD tensor $D_{ab}$ shape can be analyzed by considering the corresponding point group symmetry~\cite{Sodemann2015}.
For instance, TaAs-type compounds belong to the $C_{4v}$ point group, where $\mathcal{M}_x$ and $\mathcal{M}_y$ reflection symmetries exist.
Because $\Omega_x$ and $k_x$ are even and odd, respectively, to $\mathcal{M}_x$, $d_{xx}$ is odd to $\mathcal{M}_x$, so $D_{xx}=0$. Similarly, $D_{yy}=D_{zz}=0$. Because $\Omega_z$ and $k_x$ are odd and even, respectively, to $\mathcal{M}_y$, $d_{xz}$ is odd to $\mathcal{M}_y$, so $D_{xz}=0$.
Likewise, we obtain only two nonzero tensor elements $D_{xy}$ and $D_{yx}$, which follow $D_{xy} = -D_{yx}$.
For (Mo, W)Te$_2$ in the $C_{2v}$ point group, we obtain two nonzero independent tenor elements: $D_{xy}$ and $D_{yx}$.

Because it is a Fermi surface property, the BCD relies on the Fermi energy in the band structure.
As shown in Figs. 2a and 2b, $D_{xy}$ of TaAs exhibits a sensitive dependence on the Fermi energy. Two groups of type-I Weyl points are known to exist owing to the crossings between the top valence and bottom conduction bands: four pairs of Weyl points, noted as W1 in the $k_z=0$ plane; and eight pairs of Weyl points, noted as W2 in the $k_z\approx \pi/c$ plane ($c$ is the lattice parameter along the $z$ axis).
W1 and W2 lie 23 and 14 meV, respectively, below the charge neutral point ($E_F = 0$) (see Fig. 2b).
This is consistent with previous calculations and experimental measurements~\cite{Arnold2016FS}.
$D_{xy}$ shows a peak in magnitude when $E_F$ is close to W1, while it reverses the sign without a clear peak when $E_F$ approaches W2.
Although $D_{xy}$ is zero as $E_F$ exactly meets the Weyl point, the induced small $D_{xy}$ region can be very narrow compared to the energy sampling interval (0.1 meV in Fig. 2a).
Thus, $D_{xy}$ does not necessarily show an apparent dip of amplitude at W1 or W2.
Because $E_F=0$ is slightly away from the Weyl points, $D_{xy}$ here is smaller in magnitude than those near W1 or W2.
Fig. 2d plots $d_{xy}$ projected to the $k_x k_y$ plane. It is clear that $d_{xy}$ is mainly distributed in the W1 and W2 regions near the $\mathcal{M}_x$ plane but not the $\mathcal{M}_y$ mirror plane. Note that $d_{xy}$ does not necessarily follow the $C_4$ rotational symmetry.
When the vicinity of W1 or W2 is magnified, a roughly $xy$-like distribution and ellipse-like hollow region can be observed. Such a hollow region corresponds to the Fermi surface that surrounds a Weyl point. This is similar to the effective deduction of the Fermi surface of a tilted Weyl cone, as demonstrated in Fig. 1b.
Another striking feature is the large peak of $D_{xy}$ at $E_F=75 $ meV.
At this energy, we actually observed eight pairs of new Weyl points (noted as W3) by the crossings between the lowest and second-lowest conduction bands (see Fig. 2b). The W3 Weyl points are located between W1 and W2 in the momentum space and belong to type-II, as revealed by their energy dispersions.
The corresponding $d_{xy}$ indeed presents hot spots near W3, similar to that shown in Fig. 1c.
This further confirms that type-II Weyl points contribute a larger BCD than type-I Weyl points under similar material conditions.

\begin{figure}[htbp]
\begin{center}
\includegraphics[width=0.5\textwidth]{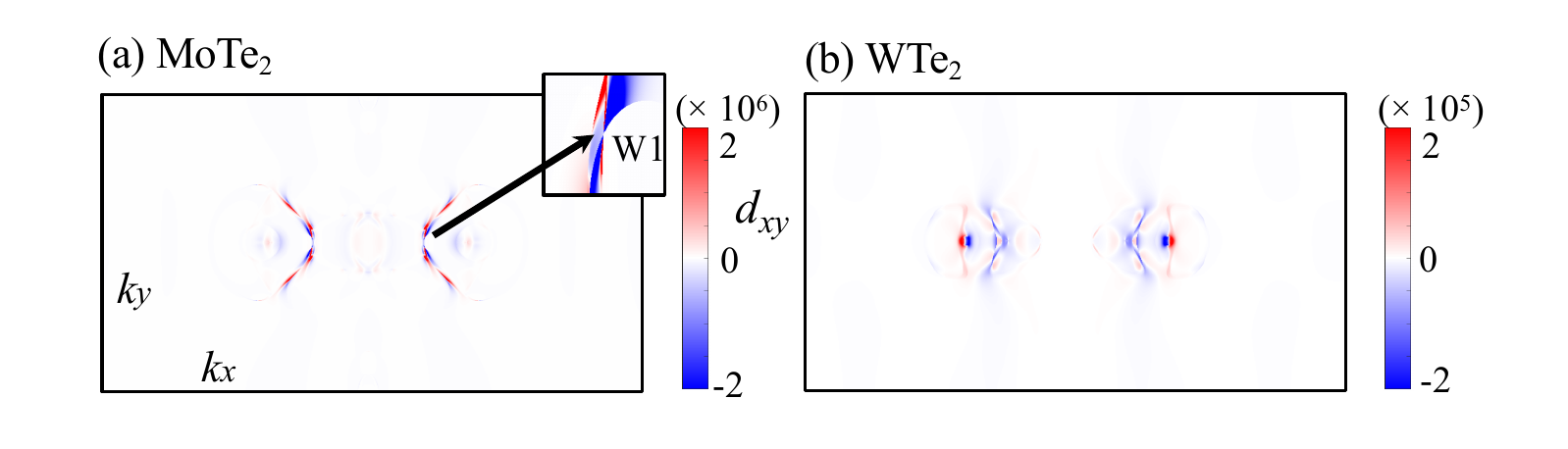}
\end{center}
\caption{
Berry curvature dipole for (a) MoTe$_2$ and (b) WTe$_2$ projected to the $k_x k_y$ plane for $E_F =0$.
A Weyl point region is magnified to demonstrate the type-II Weyl point feature.
We point out that the scale of the colorbar in (a) is one order of magnitude larger than that in (b).
}
\end{figure}

We now turn to the type-II WSMs: MoTe$_2$ and WTe$_2$.
In band structures, we obtained two pairs of type-II Weyl points for MoTe$_2$ and no Weyl point for WTe$_2$ between their conduction and valence bands, which is slightly different from the  literature~\cite{Soluyanov2015WTe2,Sun2015MoTe2}.
This discrepancy is caused by the subtle difference between different DFT methods, as revealed in recent calculations~\cite{Wang2016MoTe2,Bruno2016}.
Here, WTe$_2$ serves an example of a non-WSM for the purpose of comparison to a WSM.
For MoTe$_2$, we labeled the Weyl points as W1.
For W1 points located nearly at $E_F = 0$, $D_{xy}$ indeed shows a peak here.
Near W1, the profile of $d_{xy}$ looks like two crossing lines, which is a typical feature of the type-II Weyl point (see Fig. 3a).
In contrast, WTe$_2$ exhibits a much smaller $D_{xy}$ than MoTe$_2$.
Although some hot spots of $d_{xy}$ appear in Fig. 3b, they are less focused and one order of magnitude smaller than those of MoTe$_2$.

\begin{table}[htp]
\caption{The Berry curvature dipole calculated for Weyl semimetal materials.
The Fermi energy is set to the charge neutral point.
Only the nonzero tensor elements are listed for a given material, which are dimensionless.
}
\begin{center}
\begin{tabular}{lr|lrr}
\hline
Material & $D_{xy}$ & Material & $D_{xy}$ & $D_{yx}$ \\
\hline
TaAs & 0.39 & MoTe$_2$ & 0.849 & -0.703 \\
TaP & 0.029 & WTe$_2$ & 0.048 & -0.066 \\
NbAs &-9.88 & & & \\
NbP &20.06 & & & \\
\hline
\end{tabular}
\end{center}
\label{default}
\end{table}%

\textit{Discussion --}
Based on the results for TaAs, MoTe$_2$ and WTe$_2$, we verified these features of BCD as observed in simple models. Weyl points induce a large BCD, and type-II Weyl points usually contribute much more than type-I Weyl points. A WSM possibly exhibits a much stronger nonlinear response than an ordinary metal.

Furthermore, we reveal some new features of a BCD when all six compounds in Table I are compared.
It is known that TaAs, TaP, NbAs, and NbP exhibit a decreasing order of spin-orbit coupling (SOC), which leads to a similarly decreasing order of the spin Hall effect~\cite{Sun2016}.
However, $D_{xy}$ does not follow the same order of SOC.
NbAs and NbP show a much larger $D_{xy}$ than other materials, including MoTe$_2$.

\begin{figure}[htbp]
\begin{center}
\includegraphics[width=0.5\textwidth]{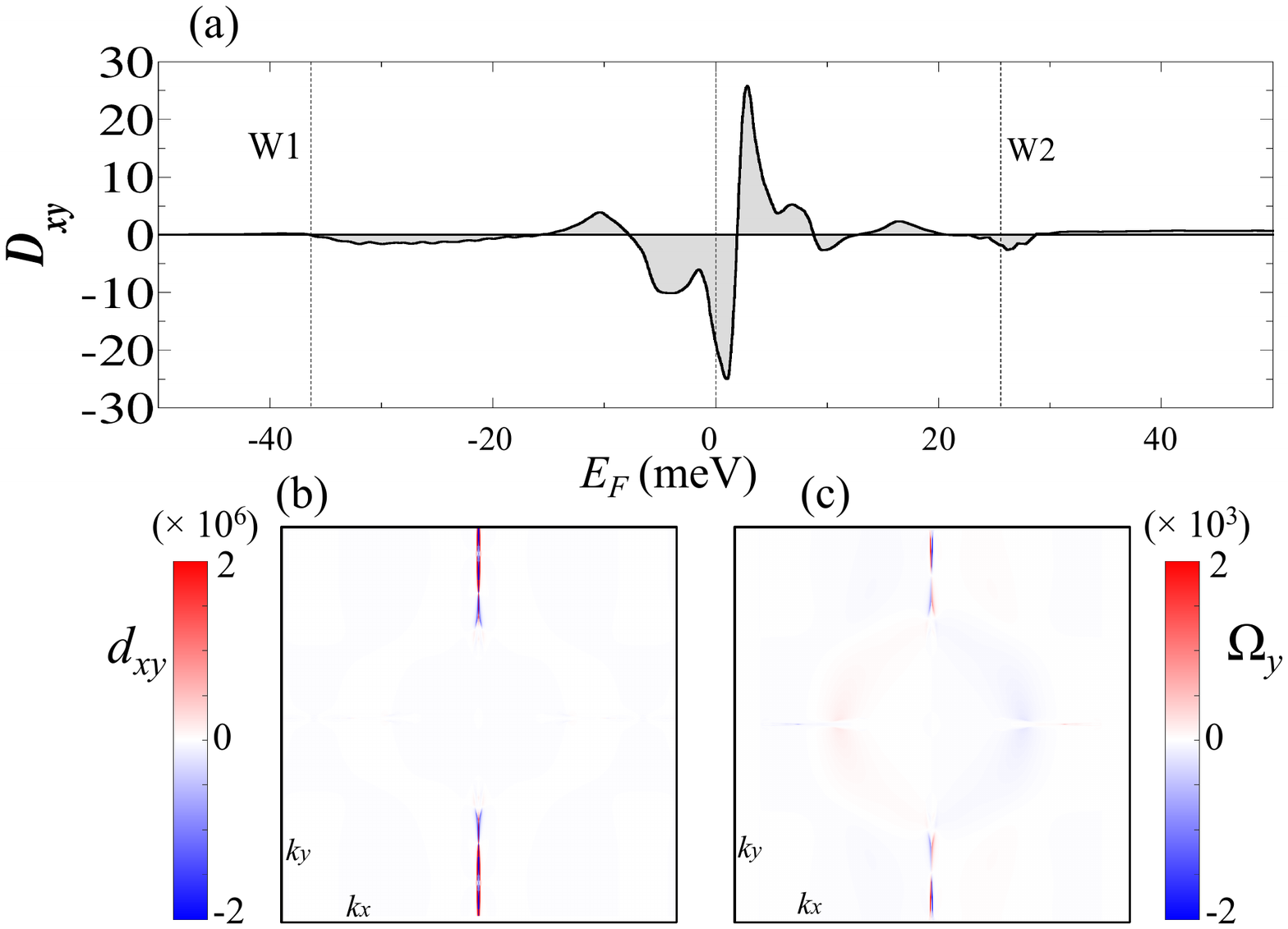}
\end{center}
\caption{Berry curvature dipole for NbP for $E_F =0$.
(a) The energy dependent $D_{xy}$. W1 and W2 Weyl points lie below and above the charge neutral point, respectively, as indicated.
(b) The $d_{xy}$ and (c) $\Omega_y$ distributions at $E_F = 0$.
Both are shown in arbitrary units.
}
\end{figure}

Take NbP as an example.
Its W1 Weyl points (36 meV below $E_F=0$) present rather small $D_{xy}$ because they are type-I with a weak tilt.
In contrast, its W2 points (26 meV above $E_F=0$) contribute a peak of $D_{xy}$, as shown in Fig. 4a, because they are type-II as revealed recently~\cite{Wu2017}.
Although W1 and W2 still fit the above understanding about Weyl points, there are two large peaks of $D_{xy}$ near $E_F=0$ where only trivial Femi pockets exist~\cite{Klotz2016}, which implies unusual $D_{xy}$ contributions beyond Weyl points.
We found that $d_{xy}$ is dominantly distributed along the $\mathcal{M}_x$ mirror plane (Fig. 4b).
This can be rationalized by the distribution of the Berry curvature $\Omega_y$.
$\Omega_y$ is odd to $\mathcal{M}_x$ and even to $\mathcal{M}_y$.
Therefore, the $\Omega_y$ gradient along $k_x$, i.e., $d_{xy} = \partial{\Omega_y}/ \partial{k_x}$, is large when crossing the $\mathcal{M}_x$ plane.
Further, $\Omega_y$ is more concentrated in a small region very close to the $\mathcal{M}_x$ plane in NbP than in TaAs, which further enhances $d_{xy}$ in NbP.
Such a difference between NbP and TaAs originates from their band structures of SOC.
At the limit of zero SOC, the valence and conduction bands each cross inside a mirror plane, which gives rise to a mirror symmetry protected nodal ring for the two systems.
As the SOC increases, the nodal ring is gapped out.
Consequently, the Berry curvature that caused by the entanglement between the valence and conduction bands has a more extended distribution in momentum space, as indicated by Eq.~\ref{eq:berry}.
Therefore, TaAs, with a stronger SOC, exhibits a much smaller BCD than NbP.

The sensitive Fermi surface dependence of the BCD serves as a way to effectively tune the nonlinear response in WSM materials. For example, both the carrier doping and external pressure are known to engineer the Fermi surface of these compounds (e.g., Refs.~\onlinecite{Sun2015MoTe2,Wang2016MoTe2,Reis2016,Belopolski2016,Einaga2017}).

Finally, we developed a semi-quantitative estimation of the nonlinear response for these materials.
According to Eq.~\ref{eq:berry}, $D_{xy}$ corresponds to $\chi_{zxx}$ and $\chi_{xxz}$, and $D_{yx}$ corresponds to $\chi_{zyy}$ and $\chi_{yyz}$.
We considered the nonlinear Hall effect, where the transverse Hall-like current $j_z = 2\chi_{zxx} \mathcal{E}_x^2$. The longitudinal current is $j_x = \sigma_{xx} \mathcal{E}_x$, and $\sigma_{xx}$ is the ordinary conductivity.
To characterize the strength of the nonlinear Hall effect, we can define the Hall angle as $\gamma = j_z / j_x = 2 (\chi_{zxx}/\sigma_{xx}) \mathcal{E}_x$.
It is known that $\gamma \sim10^{-3}$ for materials with the usual anomalous Hall effect (see Ref.~\onlinecite{Nagaosa2010} for a review).
We assumed the relaxation time $\uptau \sim 10 $ ps and $\sigma_{xx} \sim 10^6 ~\mathrm{\Omega^{-1} m^{-1}}$ based on recent low-temperature experiments (e.g., Refs.~\onlinecite{Shekhar2015,Arnold2016,Zhang2016ABJ,Huang2015anomaly,Qi2016MoTe2}) and an electric field $\mathcal{E}_x \sim10^2 ~\mathrm{V/m}$ that is applicable for a physical property measurement system in a laboratory.
Then, we obtained $\chi_{zxx} \sim 10^{-1} D_{xy}$ and $\gamma \sim 10^{-5} - 10^{-4}$ for $D_{xy}$ in the range of TaAs, MoTe$_2$, NbAs and NbP.
Because $\gamma$ of the nonlinear Hall effect approaches 10\% of that of anomalous Hall systems, the nonlinear Hall effect can be measurable for these WSM compounds.

\textit{Acknowledgments -- }
We thank Prof. Claudia Felser, Prof. Jeroen van den Brink,  Prof. Charles Kane, Prof. Eugene Mele, and Dr. Inti Sodemann for the helpful discussions. Y.Z. thanks financial support by the German Research Foundation (DFG, SFB 1143). B.Y. acknowledges the Ruth and Herman Albert Scholars Program for New Scientists of the Weizmann Institute of Science, Israel, and a grant from the German-Israeli Foundation for Scientific Research and Development (GIF Grant no. I-1364-303.7/2016).


\begin{thebibliography}{56}%
\makeatletter
\providecommand \@ifxundefined [1]{%
 \@ifx{#1\undefined}
}%
\providecommand \@ifnum [1]{%
 \ifnum #1\expandafter \@firstoftwo
 \else \expandafter \@secondoftwo
 \fi
}%
\providecommand \@ifx [1]{%
 \ifx #1\expandafter \@firstoftwo
 \else \expandafter \@secondoftwo
 \fi
}%
\providecommand \natexlab [1]{#1}%
\providecommand \enquote  [1]{``#1''}%
\providecommand \bibnamefont  [1]{#1}%
\providecommand \bibfnamefont [1]{#1}%
\providecommand \citenamefont [1]{#1}%
\providecommand \href@noop [0]{\@secondoftwo}%
\providecommand \href [0]{\begingroup \@sanitize@url \@href}%
\providecommand \@href[1]{\@@startlink{#1}\@@href}%
\providecommand \@@href[1]{\endgroup#1\@@endlink}%
\providecommand \@sanitize@url [0]{\catcode `\\12\catcode `\$12\catcode
  `\&12\catcode `\#12\catcode `\^12\catcode `\_12\catcode `\%12\relax}%
\providecommand \@@startlink[1]{}%
\providecommand \@@endlink[0]{}%
\providecommand \url  [0]{\begingroup\@sanitize@url \@url }%
\providecommand \@url [1]{\endgroup\@href {#1}{\urlprefix }}%
\providecommand \urlprefix  [0]{URL }%
\providecommand \Eprint [0]{\href }%
\providecommand \doibase [0]{http://dx.doi.org/}%
\providecommand \selectlanguage [0]{\@gobble}%
\providecommand \bibinfo  [0]{\@secondoftwo}%
\providecommand \bibfield  [0]{\@secondoftwo}%
\providecommand \translation [1]{[#1]}%
\providecommand \BibitemOpen [0]{}%
\providecommand \bibitemStop [0]{}%
\providecommand \bibitemNoStop [0]{.\EOS\space}%
\providecommand \EOS [0]{\spacefactor3000\relax}%
\providecommand \BibitemShut  [1]{\csname bibitem#1\endcsname}%
\let\auto@bib@innerbib\@empty
\bibitem [{\citenamefont {Wan}\ \emph {et~al.}(2011)\citenamefont {Wan},
  \citenamefont {Turner}, \citenamefont {Vishwanath},\ and\ \citenamefont
  {Savrasov}}]{Wan2011}%
  \BibitemOpen
  \bibfield  {author} {\bibinfo {author} {\bibfnamefont {X.~G.}\ \bibnamefont
  {Wan}}, \bibinfo {author} {\bibfnamefont {A.~M.}\ \bibnamefont {Turner}},
  \bibinfo {author} {\bibfnamefont {A.}~\bibnamefont {Vishwanath}}, \ and\
  \bibinfo {author} {\bibfnamefont {S.~Y.}\ \bibnamefont {Savrasov}},\
  }\href@noop {} {\bibfield  {journal} {\bibinfo  {journal} {Phys. Rev. B}\
  }\textbf {\bibinfo {volume} {83}},\ \bibinfo {pages} {205101} (\bibinfo
  {year} {2011})}\BibitemShut {NoStop}%
\bibitem [{\citenamefont {Volovik}(2003)}]{volovik2003universe}%
  \BibitemOpen
  \bibfield  {author} {\bibinfo {author} {\bibfnamefont {G.~E.}\ \bibnamefont
  {Volovik}},\ }\href@noop {} {\emph {\bibinfo {title} {{The Universe in A
  Helium Droplet}}}}\ (\bibinfo  {publisher} {Clarendon Press, Oxford},\
  \bibinfo {year} {2003})\BibitemShut {NoStop}%
\bibitem [{\citenamefont {Burkov}\ \emph {et~al.}(2011)\citenamefont {Burkov},
  \citenamefont {Hook},\ and\ \citenamefont {Balents}}]{Burkov2011}%
  \BibitemOpen
  \bibfield  {author} {\bibinfo {author} {\bibfnamefont {A.~A.}\ \bibnamefont
  {Burkov}}, \bibinfo {author} {\bibfnamefont {M.~D.}\ \bibnamefont {Hook}}, \
  and\ \bibinfo {author} {\bibfnamefont {L.}~\bibnamefont {Balents}},\
  }\href@noop {} {\bibfield  {journal} {\bibinfo  {journal} {Phys. Rev. B}\
  }\textbf {\bibinfo {volume} {84}},\ \bibinfo {pages} {235126} (\bibinfo
  {year} {2011})}\BibitemShut {NoStop}%
\bibitem [{\citenamefont {Hosur}\ and\ \citenamefont {Qi}(2013)}]{Hosur2013}%
  \BibitemOpen
  \bibfield  {author} {\bibinfo {author} {\bibfnamefont {P.}~\bibnamefont
  {Hosur}}\ and\ \bibinfo {author} {\bibfnamefont {X.~L.}\ \bibnamefont {Qi}},\
  }\href@noop {} {\bibfield  {journal} {\bibinfo  {journal} {C. R. Physique}\
  }\textbf {\bibinfo {volume} {14}},\ \bibinfo {pages} {857} (\bibinfo {year}
  {2013})}\BibitemShut {NoStop}%
\bibitem [{\citenamefont {Yan}\ and\ \citenamefont {Felser}(2017)}]{Yan2017}%
  \BibitemOpen
  \bibfield  {author} {\bibinfo {author} {\bibfnamefont {B.}~\bibnamefont
  {Yan}}\ and\ \bibinfo {author} {\bibfnamefont {C.}~\bibnamefont {Felser}},\
  }\href@noop {} {\bibfield  {journal} {\bibinfo  {journal} {Annual Review of
  Condensed Matter Physics}\ }\textbf {\bibinfo {volume} {8}},\ \bibinfo
  {pages} {337} (\bibinfo {year} {2017})}\BibitemShut {NoStop}%
\bibitem [{\citenamefont {Armitage}\ \emph {et~al.}(2017)\citenamefont
  {Armitage}, \citenamefont {Mele},\ and\ \citenamefont
  {Vishwanath}}]{Armitage2017}%
  \BibitemOpen
  \bibfield  {author} {\bibinfo {author} {\bibfnamefont {N.~P.}\ \bibnamefont
  {Armitage}}, \bibinfo {author} {\bibfnamefont {E.~J.}\ \bibnamefont {Mele}},
  \ and\ \bibinfo {author} {\bibfnamefont {A.}~\bibnamefont {Vishwanath}},\
  }\href@noop {} {\bibfield  {journal} {\bibinfo  {journal} {arxiv}\ }
  (\bibinfo {year} {2017})},\ \Eprint {http://arxiv.org/abs/1705.01111}
  {1705.01111} \BibitemShut {NoStop}%
\bibitem [{\citenamefont {Weng}\ \emph {et~al.}(2015)\citenamefont {Weng},
  \citenamefont {Fang}, \citenamefont {Fang}, \citenamefont {Bernevig},\ and\
  \citenamefont {Dai}}]{Weng2015}%
  \BibitemOpen
  \bibfield  {author} {\bibinfo {author} {\bibfnamefont {H.}~\bibnamefont
  {Weng}}, \bibinfo {author} {\bibfnamefont {C.}~\bibnamefont {Fang}}, \bibinfo
  {author} {\bibfnamefont {Z.}~\bibnamefont {Fang}}, \bibinfo {author}
  {\bibfnamefont {B.~A.}\ \bibnamefont {Bernevig}}, \ and\ \bibinfo {author}
  {\bibfnamefont {X.}~\bibnamefont {Dai}},\ }\href@noop {} {\bibfield
  {journal} {\bibinfo  {journal} {Phys. Rev. X}\ }\textbf {\bibinfo {volume}
  {5}},\ \bibinfo {pages} {011029} (\bibinfo {year} {2015})}\BibitemShut
  {NoStop}%
\bibitem [{\citenamefont {Huang}\ \emph
  {et~al.}(2015{\natexlab{a}})\citenamefont {Huang}, \citenamefont {Xu},
  \citenamefont {Belopolski}, \citenamefont {Lee}, \citenamefont {Chang},
  \citenamefont {Wang}, \citenamefont {Alidoust}, \citenamefont {Bian},
  \citenamefont {Neupane}, \citenamefont {Zhang}, \citenamefont {Jia},
  \citenamefont {Bansil}, \citenamefont {Lin},\ and\ \citenamefont
  {Hasan}}]{Huang2015}%
  \BibitemOpen
  \bibfield  {author} {\bibinfo {author} {\bibfnamefont {S.-M.}\ \bibnamefont
  {Huang}}, \bibinfo {author} {\bibfnamefont {S.-Y.}\ \bibnamefont {Xu}},
  \bibinfo {author} {\bibfnamefont {I.}~\bibnamefont {Belopolski}}, \bibinfo
  {author} {\bibfnamefont {C.-C.}\ \bibnamefont {Lee}}, \bibinfo {author}
  {\bibfnamefont {G.}~\bibnamefont {Chang}}, \bibinfo {author} {\bibfnamefont
  {B.}~\bibnamefont {Wang}}, \bibinfo {author} {\bibfnamefont {N.}~\bibnamefont
  {Alidoust}}, \bibinfo {author} {\bibfnamefont {G.}~\bibnamefont {Bian}},
  \bibinfo {author} {\bibfnamefont {M.}~\bibnamefont {Neupane}}, \bibinfo
  {author} {\bibfnamefont {C.}~\bibnamefont {Zhang}}, \bibinfo {author}
  {\bibfnamefont {S.}~\bibnamefont {Jia}}, \bibinfo {author} {\bibfnamefont
  {A.}~\bibnamefont {Bansil}}, \bibinfo {author} {\bibfnamefont
  {H.}~\bibnamefont {Lin}}, \ and\ \bibinfo {author} {\bibfnamefont {M.~Z.}\
  \bibnamefont {Hasan}},\ }\href@noop {} {\bibfield  {journal} {\bibinfo
  {journal} {Nat. Commun.}\ }\textbf {\bibinfo {volume} {6}},\ \bibinfo {pages}
  {8373} (\bibinfo {year} {2015}{\natexlab{a}})}\BibitemShut {NoStop}%
\bibitem [{\citenamefont {Soluyanov}\ \emph {et~al.}(2015)\citenamefont
  {Soluyanov}, \citenamefont {Gresch}, \citenamefont {Wang}, \citenamefont
  {Wu}, \citenamefont {Troyer}, \citenamefont {Dai},\ and\ \citenamefont
  {Bernevig}}]{Soluyanov2015WTe2}%
  \BibitemOpen
  \bibfield  {author} {\bibinfo {author} {\bibfnamefont {A.~A.}\ \bibnamefont
  {Soluyanov}}, \bibinfo {author} {\bibfnamefont {D.}~\bibnamefont {Gresch}},
  \bibinfo {author} {\bibfnamefont {Z.}~\bibnamefont {Wang}}, \bibinfo {author}
  {\bibfnamefont {Q.}~\bibnamefont {Wu}}, \bibinfo {author} {\bibfnamefont
  {M.}~\bibnamefont {Troyer}}, \bibinfo {author} {\bibfnamefont
  {X.}~\bibnamefont {Dai}}, \ and\ \bibinfo {author} {\bibfnamefont {B.~A.}\
  \bibnamefont {Bernevig}},\ }\href@noop {} {\bibfield  {journal} {\bibinfo
  {journal} {Nature}\ }\textbf {\bibinfo {volume} {527}},\ \bibinfo {pages}
  {495} (\bibinfo {year} {2015})}\BibitemShut {NoStop}%
\bibitem [{\citenamefont {Sun}\ \emph {et~al.}(2015)\citenamefont {Sun},
  \citenamefont {Wu}, \citenamefont {Ali}, \citenamefont {Felser},\ and\
  \citenamefont {Yan}}]{Sun2015MoTe2}%
  \BibitemOpen
  \bibfield  {author} {\bibinfo {author} {\bibfnamefont {Y.}~\bibnamefont
  {Sun}}, \bibinfo {author} {\bibfnamefont {S.~C.}\ \bibnamefont {Wu}},
  \bibinfo {author} {\bibfnamefont {M.~N.}\ \bibnamefont {Ali}}, \bibinfo
  {author} {\bibfnamefont {C.}~\bibnamefont {Felser}}, \ and\ \bibinfo {author}
  {\bibfnamefont {B.}~\bibnamefont {Yan}},\ }\href@noop {} {\bibfield
  {journal} {\bibinfo  {journal} {Phys. Rev. B}\ }\textbf {\bibinfo {volume}
  {92}},\ \bibinfo {pages} {161107(R)} (\bibinfo {year} {2015})}\BibitemShut
  {NoStop}%
\bibitem [{\citenamefont {Lv}\ \emph {et~al.}(2015)\citenamefont {Lv},
  \citenamefont {Weng}, \citenamefont {Fu}, \citenamefont {Wang}, \citenamefont
  {Miao}, \citenamefont {Ma}, \citenamefont {Richard}, \citenamefont {Huang},
  \citenamefont {Zhao}, \citenamefont {Chen}, \citenamefont {Fang},
  \citenamefont {Dai}, \citenamefont {Qian},\ and\ \citenamefont
  {Ding}}]{Lv2015TaAs}%
  \BibitemOpen
  \bibfield  {author} {\bibinfo {author} {\bibfnamefont {B.~Q.}\ \bibnamefont
  {Lv}}, \bibinfo {author} {\bibfnamefont {H.~M.}\ \bibnamefont {Weng}},
  \bibinfo {author} {\bibfnamefont {B.~B.}\ \bibnamefont {Fu}}, \bibinfo
  {author} {\bibfnamefont {X.~P.}\ \bibnamefont {Wang}}, \bibinfo {author}
  {\bibfnamefont {H.}~\bibnamefont {Miao}}, \bibinfo {author} {\bibfnamefont
  {J.}~\bibnamefont {Ma}}, \bibinfo {author} {\bibfnamefont {P.}~\bibnamefont
  {Richard}}, \bibinfo {author} {\bibfnamefont {X.~C.}\ \bibnamefont {Huang}},
  \bibinfo {author} {\bibfnamefont {L.~X.}\ \bibnamefont {Zhao}}, \bibinfo
  {author} {\bibfnamefont {G.~F.}\ \bibnamefont {Chen}}, \bibinfo {author}
  {\bibfnamefont {Z.}~\bibnamefont {Fang}}, \bibinfo {author} {\bibfnamefont
  {X.}~\bibnamefont {Dai}}, \bibinfo {author} {\bibfnamefont {T.}~\bibnamefont
  {Qian}}, \ and\ \bibinfo {author} {\bibfnamefont {H.}~\bibnamefont {Ding}},\
  }\href@noop {} {\bibfield  {journal} {\bibinfo  {journal} {Phys. Rev. X}\
  }\textbf {\bibinfo {volume} {5}},\ \bibinfo {pages} {031013} (\bibinfo {year}
  {2015})}\BibitemShut {NoStop}%
\bibitem [{\citenamefont {Xu}\ \emph {et~al.}(2015)\citenamefont {Xu},
  \citenamefont {Belopolski}, \citenamefont {Alidoust}, \citenamefont
  {Neupane}, \citenamefont {Bian}, \citenamefont {Zhang}, \citenamefont
  {Sankar}, \citenamefont {Chang}, \citenamefont {Zhujun}, \citenamefont {Lee},
  \citenamefont {Shin-Ming}, \citenamefont {Zheng}, \citenamefont {Ma},
  \citenamefont {Sanchez}, \citenamefont {Wang}, \citenamefont {Bansil},
  \citenamefont {Chou}, \citenamefont {Shibayev}, \citenamefont {Lin},
  \citenamefont {Jia},\ and\ \citenamefont {Hasan}}]{Xu2015TaAs}%
  \BibitemOpen
  \bibfield  {author} {\bibinfo {author} {\bibfnamefont {S.-Y.}\ \bibnamefont
  {Xu}}, \bibinfo {author} {\bibfnamefont {I.}~\bibnamefont {Belopolski}},
  \bibinfo {author} {\bibfnamefont {N.}~\bibnamefont {Alidoust}}, \bibinfo
  {author} {\bibfnamefont {M.}~\bibnamefont {Neupane}}, \bibinfo {author}
  {\bibfnamefont {G.}~\bibnamefont {Bian}}, \bibinfo {author} {\bibfnamefont
  {C.}~\bibnamefont {Zhang}}, \bibinfo {author} {\bibfnamefont
  {R.}~\bibnamefont {Sankar}}, \bibinfo {author} {\bibfnamefont
  {G.}~\bibnamefont {Chang}}, \bibinfo {author} {\bibfnamefont
  {Y.}~\bibnamefont {Zhujun}}, \bibinfo {author} {\bibfnamefont {C.-C.}\
  \bibnamefont {Lee}}, \bibinfo {author} {\bibfnamefont {H.}~\bibnamefont
  {Shin-Ming}}, \bibinfo {author} {\bibfnamefont {H.}~\bibnamefont {Zheng}},
  \bibinfo {author} {\bibfnamefont {J.}~\bibnamefont {Ma}}, \bibinfo {author}
  {\bibfnamefont {D.~S.}\ \bibnamefont {Sanchez}}, \bibinfo {author}
  {\bibfnamefont {B.}~\bibnamefont {Wang}}, \bibinfo {author} {\bibfnamefont
  {A.}~\bibnamefont {Bansil}}, \bibinfo {author} {\bibfnamefont
  {F.}~\bibnamefont {Chou}}, \bibinfo {author} {\bibfnamefont {P.~P.}\
  \bibnamefont {Shibayev}}, \bibinfo {author} {\bibfnamefont {H.}~\bibnamefont
  {Lin}}, \bibinfo {author} {\bibfnamefont {S.}~\bibnamefont {Jia}}, \ and\
  \bibinfo {author} {\bibfnamefont {M.~Z.}\ \bibnamefont {Hasan}},\ }\href@noop
  {} {\bibfield  {journal} {\bibinfo  {journal} {Science}\ }\textbf {\bibinfo
  {volume} {349}},\ \bibinfo {pages} {613} (\bibinfo {year}
  {2015})}\BibitemShut {NoStop}%
\bibitem [{\citenamefont {Yang}\ \emph {et~al.}(2015)\citenamefont {Yang},
  \citenamefont {Liu}, \citenamefont {Sun}, \citenamefont {Peng}, \citenamefont
  {Yang}, \citenamefont {Zhang}, \citenamefont {Zhou}, \citenamefont {Zhang},
  \citenamefont {Guo}, \citenamefont {Rahn}, \citenamefont {Prabhakaran},
  \citenamefont {Hussain}, \citenamefont {Mo}, \citenamefont {Felser},
  \citenamefont {Yan},\ and\ \citenamefont {Chen}}]{Yang2015TaAs}%
  \BibitemOpen
  \bibfield  {author} {\bibinfo {author} {\bibfnamefont {L.~X.}\ \bibnamefont
  {Yang}}, \bibinfo {author} {\bibfnamefont {Z.~K.}\ \bibnamefont {Liu}},
  \bibinfo {author} {\bibfnamefont {Y.}~\bibnamefont {Sun}}, \bibinfo {author}
  {\bibfnamefont {H.}~\bibnamefont {Peng}}, \bibinfo {author} {\bibfnamefont
  {H.~F.}\ \bibnamefont {Yang}}, \bibinfo {author} {\bibfnamefont
  {T.}~\bibnamefont {Zhang}}, \bibinfo {author} {\bibfnamefont
  {B.}~\bibnamefont {Zhou}}, \bibinfo {author} {\bibfnamefont {Y.}~\bibnamefont
  {Zhang}}, \bibinfo {author} {\bibfnamefont {Y.~F.}\ \bibnamefont {Guo}},
  \bibinfo {author} {\bibfnamefont {M.}~\bibnamefont {Rahn}}, \bibinfo {author}
  {\bibfnamefont {D.}~\bibnamefont {Prabhakaran}}, \bibinfo {author}
  {\bibfnamefont {Z.}~\bibnamefont {Hussain}}, \bibinfo {author} {\bibfnamefont
  {S.~K.}\ \bibnamefont {Mo}}, \bibinfo {author} {\bibfnamefont
  {C.}~\bibnamefont {Felser}}, \bibinfo {author} {\bibfnamefont
  {B.}~\bibnamefont {Yan}}, \ and\ \bibinfo {author} {\bibfnamefont {Y.~L.}\
  \bibnamefont {Chen}},\ }\href@noop {} {\bibfield  {journal} {\bibinfo
  {journal} {Nat. Phys.}\ }\textbf {\bibinfo {volume} {11}},\ \bibinfo {pages}
  {728} (\bibinfo {year} {2015})}\BibitemShut {NoStop}%
\bibitem [{\citenamefont {Deng}\ \emph {et~al.}(2016)\citenamefont {Deng},
  \citenamefont {Wan}, \citenamefont {Deng}, \citenamefont {Zhang},
  \citenamefont {Ding}, \citenamefont {Wang}, \citenamefont {Yan},
  \citenamefont {Huang}, \citenamefont {Zhang}, \citenamefont {Xu},
  \citenamefont {Denlinger}, \citenamefont {Fedorov}, \citenamefont {Yang},
  \citenamefont {Duan}, \citenamefont {Yao}, \citenamefont {Wu}, \citenamefont
  {Fan}, \citenamefont {Zhang}, \citenamefont {Chen},\ and\ \citenamefont
  {Zhou}}]{Deng2016}%
  \BibitemOpen
  \bibfield  {author} {\bibinfo {author} {\bibfnamefont {K.}~\bibnamefont
  {Deng}}, \bibinfo {author} {\bibfnamefont {G.}~\bibnamefont {Wan}}, \bibinfo
  {author} {\bibfnamefont {P.}~\bibnamefont {Deng}}, \bibinfo {author}
  {\bibfnamefont {K.}~\bibnamefont {Zhang}}, \bibinfo {author} {\bibfnamefont
  {S.}~\bibnamefont {Ding}}, \bibinfo {author} {\bibfnamefont {E.}~\bibnamefont
  {Wang}}, \bibinfo {author} {\bibfnamefont {M.}~\bibnamefont {Yan}}, \bibinfo
  {author} {\bibfnamefont {H.}~\bibnamefont {Huang}}, \bibinfo {author}
  {\bibfnamefont {H.}~\bibnamefont {Zhang}}, \bibinfo {author} {\bibfnamefont
  {Z.}~\bibnamefont {Xu}}, \bibinfo {author} {\bibfnamefont {J.}~\bibnamefont
  {Denlinger}}, \bibinfo {author} {\bibfnamefont {A.}~\bibnamefont {Fedorov}},
  \bibinfo {author} {\bibfnamefont {H.}~\bibnamefont {Yang}}, \bibinfo {author}
  {\bibfnamefont {W.}~\bibnamefont {Duan}}, \bibinfo {author} {\bibfnamefont
  {H.}~\bibnamefont {Yao}}, \bibinfo {author} {\bibfnamefont {Y.}~\bibnamefont
  {Wu}}, \bibinfo {author} {\bibfnamefont {S.}~\bibnamefont {Fan}}, \bibinfo
  {author} {\bibfnamefont {H.}~\bibnamefont {Zhang}}, \bibinfo {author}
  {\bibfnamefont {X.}~\bibnamefont {Chen}}, \ and\ \bibinfo {author}
  {\bibfnamefont {S.}~\bibnamefont {Zhou}},\ }\href@noop {} {\bibfield
  {journal} {\bibinfo  {journal} {Nat. Phys.}\ }\textbf {\bibinfo {volume}
  {12}},\ \bibinfo {pages} {1105} (\bibinfo {year} {2016})}\BibitemShut
  {NoStop}%
\bibitem [{\citenamefont {Jiang}\ \emph {et~al.}(2017)\citenamefont {Jiang},
  \citenamefont {Liu}, \citenamefont {Sun}, \citenamefont {Yang}, \citenamefont
  {Rajamathi}, \citenamefont {Qi}, \citenamefont {Yang}, \citenamefont {Chen},
  \citenamefont {Peng}, \citenamefont {Hwang}, \citenamefont {Sun},
  \citenamefont {Mo}, \citenamefont {Vobornik}, \citenamefont {Fujii},
  \citenamefont {Parkin}, \citenamefont {Felser}, \citenamefont {Yan},\ and\
  \citenamefont {Chen}}]{Jiang2016}%
  \BibitemOpen
  \bibfield  {author} {\bibinfo {author} {\bibfnamefont {J.}~\bibnamefont
  {Jiang}}, \bibinfo {author} {\bibfnamefont {Z.~K.}\ \bibnamefont {Liu}},
  \bibinfo {author} {\bibfnamefont {Y.}~\bibnamefont {Sun}}, \bibinfo {author}
  {\bibfnamefont {H.~F.}\ \bibnamefont {Yang}}, \bibinfo {author}
  {\bibfnamefont {C.~R.}\ \bibnamefont {Rajamathi}}, \bibinfo {author}
  {\bibfnamefont {Y.~P.}\ \bibnamefont {Qi}}, \bibinfo {author} {\bibfnamefont
  {L.~X.}\ \bibnamefont {Yang}}, \bibinfo {author} {\bibfnamefont
  {C.}~\bibnamefont {Chen}}, \bibinfo {author} {\bibfnamefont {H.}~\bibnamefont
  {Peng}}, \bibinfo {author} {\bibfnamefont {C.~C.}\ \bibnamefont {Hwang}},
  \bibinfo {author} {\bibfnamefont {S.~Z.}\ \bibnamefont {Sun}}, \bibinfo
  {author} {\bibfnamefont {S.-K.}\ \bibnamefont {Mo}}, \bibinfo {author}
  {\bibfnamefont {I.}~\bibnamefont {Vobornik}}, \bibinfo {author}
  {\bibfnamefont {J.}~\bibnamefont {Fujii}}, \bibinfo {author} {\bibfnamefont
  {S.}~\bibnamefont {Parkin}}, \bibinfo {author} {\bibfnamefont
  {C.}~\bibnamefont {Felser}}, \bibinfo {author} {\bibfnamefont
  {B.}~\bibnamefont {Yan}}, \ and\ \bibinfo {author} {\bibfnamefont {Y.~L.}\
  \bibnamefont {Chen}},\ }\href@noop {} {\bibfield  {journal} {\bibinfo
  {journal} {Nat. Commun.}\ }\textbf {\bibinfo {volume} {8}},\ \bibinfo {pages}
  {13973} (\bibinfo {year} {2017})}\BibitemShut {NoStop}%
\bibitem [{\citenamefont {Huang}\ \emph {et~al.}(2016)\citenamefont {Huang},
  \citenamefont {McCormick}, \citenamefont {Ochi}, \citenamefont {Zhao},
  \citenamefont {Suzuki}, \citenamefont {Arita}, \citenamefont {Wu},
  \citenamefont {Mou}, \citenamefont {Cao}, \citenamefont {Yan}, \citenamefont
  {Trivedi},\ and\ \citenamefont {Kaminski}}]{Huang2016}%
  \BibitemOpen
  \bibfield  {author} {\bibinfo {author} {\bibfnamefont {L.}~\bibnamefont
  {Huang}}, \bibinfo {author} {\bibfnamefont {T.~M.}\ \bibnamefont
  {McCormick}}, \bibinfo {author} {\bibfnamefont {M.}~\bibnamefont {Ochi}},
  \bibinfo {author} {\bibfnamefont {Z.}~\bibnamefont {Zhao}}, \bibinfo {author}
  {\bibfnamefont {M.-t.}\ \bibnamefont {Suzuki}}, \bibinfo {author}
  {\bibfnamefont {R.}~\bibnamefont {Arita}}, \bibinfo {author} {\bibfnamefont
  {Y.}~\bibnamefont {Wu}}, \bibinfo {author} {\bibfnamefont {D.}~\bibnamefont
  {Mou}}, \bibinfo {author} {\bibfnamefont {H.}~\bibnamefont {Cao}}, \bibinfo
  {author} {\bibfnamefont {J.}~\bibnamefont {Yan}}, \bibinfo {author}
  {\bibfnamefont {N.}~\bibnamefont {Trivedi}}, \ and\ \bibinfo {author}
  {\bibfnamefont {A.}~\bibnamefont {Kaminski}},\ }\href@noop {} {\bibfield
  {journal} {\bibinfo  {journal} {Nat. Mater.}\ }\textbf {\bibinfo {volume}
  {15}},\ \bibinfo {pages} {1155} (\bibinfo {year} {2016})}\BibitemShut
  {NoStop}%
\bibitem [{\citenamefont {Xu}\ \emph {et~al.}(2011)\citenamefont {Xu},
  \citenamefont {Weng}, \citenamefont {Wang}, \citenamefont {Dai},\ and\
  \citenamefont {Fang}}]{Xu2011}%
  \BibitemOpen
  \bibfield  {author} {\bibinfo {author} {\bibfnamefont {G.}~\bibnamefont
  {Xu}}, \bibinfo {author} {\bibfnamefont {H.}~\bibnamefont {Weng}}, \bibinfo
  {author} {\bibfnamefont {Z.}~\bibnamefont {Wang}}, \bibinfo {author}
  {\bibfnamefont {X.}~\bibnamefont {Dai}}, \ and\ \bibinfo {author}
  {\bibfnamefont {Z.}~\bibnamefont {Fang}},\ }\href@noop {} {\bibfield
  {journal} {\bibinfo  {journal} {{Phys. Rev. Lett.}}\ }\textbf {\bibinfo
  {volume} {107}},\ \bibinfo {pages} {186806} (\bibinfo {year}
  {2011})}\BibitemShut {NoStop}%
\bibitem [{\citenamefont {Yang}\ \emph {et~al.}(2011)\citenamefont {Yang},
  \citenamefont {Lu},\ and\ \citenamefont {Ran}}]{Yang2011QHE}%
  \BibitemOpen
  \bibfield  {author} {\bibinfo {author} {\bibfnamefont {K.-Y.}\ \bibnamefont
  {Yang}}, \bibinfo {author} {\bibfnamefont {Y.-M.}\ \bibnamefont {Lu}}, \ and\
  \bibinfo {author} {\bibfnamefont {Y.}~\bibnamefont {Ran}},\ }\href@noop {}
  {\bibfield  {journal} {\bibinfo  {journal} {Phys. Rev. B}\ }\textbf {\bibinfo
  {volume} {84}},\ \bibinfo {pages} {075129} (\bibinfo {year}
  {2011})}\BibitemShut {NoStop}%
\bibitem [{\citenamefont {Burkov}(2014)}]{Burkov2014}%
  \BibitemOpen
  \bibfield  {author} {\bibinfo {author} {\bibfnamefont {A.~A.}\ \bibnamefont
  {Burkov}},\ }\href@noop {} {\bibfield  {journal} {\bibinfo  {journal} {Phys.
  Rev. Lett.}\ }\textbf {\bibinfo {volume} {113}} (\bibinfo {year}
  {2014})}\BibitemShut {NoStop}%
\bibitem [{\citenamefont {Sun}\ \emph {et~al.}(2016{\natexlab{a}})\citenamefont
  {Sun}, \citenamefont {Zhang}, \citenamefont {Felser},\ and\ \citenamefont
  {Yan}}]{Sun2016}%
  \BibitemOpen
  \bibfield  {author} {\bibinfo {author} {\bibfnamefont {Y.}~\bibnamefont
  {Sun}}, \bibinfo {author} {\bibfnamefont {Y.}~\bibnamefont {Zhang}}, \bibinfo
  {author} {\bibfnamefont {C.}~\bibnamefont {Felser}}, \ and\ \bibinfo {author}
  {\bibfnamefont {B.}~\bibnamefont {Yan}},\ }\href@noop {} {\bibfield
  {journal} {\bibinfo  {journal} {Phys. Rev. Lett.}\ }\textbf {\bibinfo
  {volume} {117}},\ \bibinfo {pages} {146403} (\bibinfo {year}
  {2016}{\natexlab{a}})}\BibitemShut {NoStop}%
\bibitem [{\citenamefont {Hosur}\ and\ \citenamefont {Qi}(2015)}]{Hosur2015}%
  \BibitemOpen
  \bibfield  {author} {\bibinfo {author} {\bibfnamefont {P.}~\bibnamefont
  {Hosur}}\ and\ \bibinfo {author} {\bibfnamefont {X.-L.}\ \bibnamefont {Qi}},\
  }\href {\doibase 10.1103/PhysRevB.91.081106} {\bibfield  {journal} {\bibinfo
  {journal} {Phys. Rev. B}\ }\textbf {\bibinfo {volume} {91}},\ \bibinfo
  {pages} {081106} (\bibinfo {year} {2015})}\BibitemShut {NoStop}%
\bibitem [{\citenamefont {Sodemann}\ and\ \citenamefont
  {Fu}(2015)}]{Sodemann2015}%
  \BibitemOpen
  \bibfield  {author} {\bibinfo {author} {\bibfnamefont {I.}~\bibnamefont
  {Sodemann}}\ and\ \bibinfo {author} {\bibfnamefont {L.}~\bibnamefont {Fu}},\
  }\href@noop {} {\bibfield  {journal} {\bibinfo  {journal} {Phys. Rev. Lett.}\
  }\textbf {\bibinfo {volume} {115}},\ \bibinfo {pages} {216806} (\bibinfo
  {year} {2015})}\BibitemShut {NoStop}%
\bibitem [{\citenamefont {Morimoto}\ and\ \citenamefont
  {Nagaosa}(2016)}]{Morimoto2016a}%
  \BibitemOpen
  \bibfield  {author} {\bibinfo {author} {\bibfnamefont {T.}~\bibnamefont
  {Morimoto}}\ and\ \bibinfo {author} {\bibfnamefont {N.}~\bibnamefont
  {Nagaosa}},\ }\href@noop {} {\bibfield  {journal} {\bibinfo  {journal}
  {Science Advances}\ }\textbf {\bibinfo {volume} {2}},\ \bibinfo {pages}
  {e1501524} (\bibinfo {year} {2016})}\BibitemShut {NoStop}%
\bibitem [{\citenamefont {Taguchi}\ \emph {et~al.}(2016)\citenamefont
  {Taguchi}, \citenamefont {Imaeda}, \citenamefont {Sato},\ and\ \citenamefont
  {Tanaka}}]{Taguchi2016}%
  \BibitemOpen
  \bibfield  {author} {\bibinfo {author} {\bibfnamefont {K.}~\bibnamefont
  {Taguchi}}, \bibinfo {author} {\bibfnamefont {T.}~\bibnamefont {Imaeda}},
  \bibinfo {author} {\bibfnamefont {M.}~\bibnamefont {Sato}}, \ and\ \bibinfo
  {author} {\bibfnamefont {Y.}~\bibnamefont {Tanaka}},\ }\href {\doibase
  10.1103/PhysRevB.93.201202} {\bibfield  {journal} {\bibinfo  {journal} {Phys.
  Rev. B}\ }\textbf {\bibinfo {volume} {93}},\ \bibinfo {pages} {201202}
  (\bibinfo {year} {2016})}\BibitemShut {NoStop}%
\bibitem [{\citenamefont {Morimoto}\ \emph {et~al.}(2016)\citenamefont
  {Morimoto}, \citenamefont {Zhong}, \citenamefont {Orenstein},\ and\
  \citenamefont {Moore}}]{Morimoto2016}%
  \BibitemOpen
  \bibfield  {author} {\bibinfo {author} {\bibfnamefont {T.}~\bibnamefont
  {Morimoto}}, \bibinfo {author} {\bibfnamefont {S.}~\bibnamefont {Zhong}},
  \bibinfo {author} {\bibfnamefont {J.}~\bibnamefont {Orenstein}}, \ and\
  \bibinfo {author} {\bibfnamefont {J.~E.}\ \bibnamefont {Moore}},\ }\href@noop
  {} {\bibfield  {journal} {\bibinfo  {journal} {Phys. Rev. B}\ }\textbf
  {\bibinfo {volume} {94}},\ \bibinfo {pages} {245121} (\bibinfo {year}
  {2016})}\BibitemShut {NoStop}%
\bibitem [{\citenamefont {Chan}\ \emph {et~al.}(2016)\citenamefont {Chan},
  \citenamefont {Lee}, \citenamefont {Burch}, \citenamefont {Han},\ and\
  \citenamefont {Ran}}]{Chan2016}%
  \BibitemOpen
  \bibfield  {author} {\bibinfo {author} {\bibfnamefont {C.-K.}\ \bibnamefont
  {Chan}}, \bibinfo {author} {\bibfnamefont {P.~A.}\ \bibnamefont {Lee}},
  \bibinfo {author} {\bibfnamefont {K.~S.}\ \bibnamefont {Burch}}, \bibinfo
  {author} {\bibfnamefont {J.~H.}\ \bibnamefont {Han}}, \ and\ \bibinfo
  {author} {\bibfnamefont {Y.}~\bibnamefont {Ran}},\ }\href@noop {} {\bibfield
  {journal} {\bibinfo  {journal} {Phys. Rev. Lett.}\ }\textbf {\bibinfo
  {volume} {116}},\ \bibinfo {pages} {026805} (\bibinfo {year}
  {2016})}\BibitemShut {NoStop}%
\bibitem [{\citenamefont {Ishizuka}\ \emph {et~al.}(2016)\citenamefont
  {Ishizuka}, \citenamefont {Hayata}, \citenamefont {Ueda},\ and\ \citenamefont
  {Nagaosa}}]{Ishizuka2016}%
  \BibitemOpen
  \bibfield  {author} {\bibinfo {author} {\bibfnamefont {H.}~\bibnamefont
  {Ishizuka}}, \bibinfo {author} {\bibfnamefont {T.}~\bibnamefont {Hayata}},
  \bibinfo {author} {\bibfnamefont {M.}~\bibnamefont {Ueda}}, \ and\ \bibinfo
  {author} {\bibfnamefont {N.}~\bibnamefont {Nagaosa}},\ }\href {\doibase
  10.1103/PhysRevLett.117.216601} {\bibfield  {journal} {\bibinfo  {journal}
  {Phys. Rev. Lett.}\ }\textbf {\bibinfo {volume} {117}},\ \bibinfo {pages}
  {216601} (\bibinfo {year} {2016})}\BibitemShut {NoStop}%
\bibitem [{\citenamefont {de~Juan}\ \emph {et~al.}(2017)\citenamefont
  {de~Juan}, \citenamefont {Grushin}, \citenamefont {Morimoto},\ and\
  \citenamefont {Moore}}]{deJuan2017}%
  \BibitemOpen
  \bibfield  {author} {\bibinfo {author} {\bibfnamefont {F.}~\bibnamefont
  {de~Juan}}, \bibinfo {author} {\bibfnamefont {A.~G.}\ \bibnamefont
  {Grushin}}, \bibinfo {author} {\bibfnamefont {T.}~\bibnamefont {Morimoto}}, \
  and\ \bibinfo {author} {\bibfnamefont {J.~E.}\ \bibnamefont {Moore}},\
  }\href@noop {} {\bibfield  {journal} {\bibinfo  {journal} {Nat. Commun.}\
  }\textbf {\bibinfo {volume} {8}},\ \bibinfo {pages} {ncomms15995} (\bibinfo
  {year} {2017})}\BibitemShut {NoStop}%
\bibitem [{\citenamefont {Chan}\ \emph {et~al.}(2017)\citenamefont {Chan},
  \citenamefont {Lindner}, \citenamefont {Refael},\ and\ \citenamefont
  {Lee}}]{Chan2017}%
  \BibitemOpen
  \bibfield  {author} {\bibinfo {author} {\bibfnamefont {C.-K.}\ \bibnamefont
  {Chan}}, \bibinfo {author} {\bibfnamefont {N.~H.}\ \bibnamefont {Lindner}},
  \bibinfo {author} {\bibfnamefont {G.}~\bibnamefont {Refael}}, \ and\ \bibinfo
  {author} {\bibfnamefont {P.~A.}\ \bibnamefont {Lee}},\ }\href@noop {}
  {\bibfield  {journal} {\bibinfo  {journal} {Phys. Rev. B}\ }\textbf {\bibinfo
  {volume} {95}},\ \bibinfo {pages} {041104} (\bibinfo {year}
  {2017})}\BibitemShut {NoStop}%
\bibitem [{\citenamefont {Rostami}\ and\ \citenamefont
  {Polini}(2017)}]{Rostami2017}%
  \BibitemOpen
  \bibfield  {author} {\bibinfo {author} {\bibfnamefont {H.}~\bibnamefont
  {Rostami}}\ and\ \bibinfo {author} {\bibfnamefont {M.}~\bibnamefont
  {Polini}},\ }\href@noop {} {\bibfield  {journal} {\bibinfo  {journal}
  {arxiv}\ } (\bibinfo {year} {2017})},\ \Eprint
  {http://arxiv.org/abs/1705.09915} {1705.09915} \BibitemShut {NoStop}%
\bibitem [{\citenamefont {Wu}\ \emph {et~al.}(2017{\natexlab{a}})\citenamefont
  {Wu}, \citenamefont {Patankar}, \citenamefont {Morimoto}, \citenamefont
  {Nair}, \citenamefont {Thewalt}, \citenamefont {Little}, \citenamefont
  {Analytis}, \citenamefont {Moore},\ and\ \citenamefont
  {Orenstein}}]{Wu2017nonlinear}%
  \BibitemOpen
  \bibfield  {author} {\bibinfo {author} {\bibfnamefont {L.}~\bibnamefont
  {Wu}}, \bibinfo {author} {\bibfnamefont {S.}~\bibnamefont {Patankar}},
  \bibinfo {author} {\bibfnamefont {T.}~\bibnamefont {Morimoto}}, \bibinfo
  {author} {\bibfnamefont {N.~L.}\ \bibnamefont {Nair}}, \bibinfo {author}
  {\bibfnamefont {E.}~\bibnamefont {Thewalt}}, \bibinfo {author} {\bibfnamefont
  {A.}~\bibnamefont {Little}}, \bibinfo {author} {\bibfnamefont {J.~G.}\
  \bibnamefont {Analytis}}, \bibinfo {author} {\bibfnamefont {J.~E.}\
  \bibnamefont {Moore}}, \ and\ \bibinfo {author} {\bibfnamefont
  {J.}~\bibnamefont {Orenstein}},\ }\href@noop {} {\bibfield  {journal}
  {\bibinfo  {journal} {Nature Physics}\ }\textbf {\bibinfo {volume} {13}},\
  \bibinfo {pages} {350} (\bibinfo {year} {2017}{\natexlab{a}})}\BibitemShut
  {NoStop}%
\bibitem [{\citenamefont {Ma}\ \emph {et~al.}(2017)\citenamefont {Ma},
  \citenamefont {Xu}, \citenamefont {Chan}, \citenamefont {Zhang},
  \citenamefont {Chang}, \citenamefont {Lin}, \citenamefont {Xie},
  \citenamefont {Palacios}, \citenamefont {Lin}, \citenamefont {Jia},
  \citenamefont {Lee}, \citenamefont {Jarillo-Herrero},\ and\ \citenamefont
  {Gedik}}]{Ma2017}%
  \BibitemOpen
  \bibfield  {author} {\bibinfo {author} {\bibfnamefont {Q.}~\bibnamefont
  {Ma}}, \bibinfo {author} {\bibfnamefont {S.-Y.}\ \bibnamefont {Xu}}, \bibinfo
  {author} {\bibfnamefont {C.-K.}\ \bibnamefont {Chan}}, \bibinfo {author}
  {\bibfnamefont {C.-L.}\ \bibnamefont {Zhang}}, \bibinfo {author}
  {\bibfnamefont {G.}~\bibnamefont {Chang}}, \bibinfo {author} {\bibfnamefont
  {Y.}~\bibnamefont {Lin}}, \bibinfo {author} {\bibfnamefont {W.}~\bibnamefont
  {Xie}}, \bibinfo {author} {\bibfnamefont {T.}~\bibnamefont {Palacios}},
  \bibinfo {author} {\bibfnamefont {H.}~\bibnamefont {Lin}}, \bibinfo {author}
  {\bibfnamefont {S.}~\bibnamefont {Jia}}, \bibinfo {author} {\bibfnamefont
  {P.~A.}\ \bibnamefont {Lee}}, \bibinfo {author} {\bibfnamefont
  {P.}~\bibnamefont {Jarillo-Herrero}}, \ and\ \bibinfo {author} {\bibfnamefont
  {N.}~\bibnamefont {Gedik}},\ }\href@noop {} {\bibfield  {journal} {\bibinfo
  {journal} {Nature Physics}\ }\textbf {\bibinfo {volume} {56}},\ \bibinfo
  {pages} {330} (\bibinfo {year} {2017})}\BibitemShut {NoStop}%
\bibitem [{\citenamefont {Sun}\ \emph {et~al.}(2016{\natexlab{b}})\citenamefont
  {Sun}, \citenamefont {Sun}, \citenamefont {Guo}, \citenamefont {Wei},
  \citenamefont {Tian}, \citenamefont {Yang}, \citenamefont {Chen},\ and\
  \citenamefont {Li}}]{Sun2016CPGE}%
  \BibitemOpen
  \bibfield  {author} {\bibinfo {author} {\bibfnamefont {K.}~\bibnamefont
  {Sun}}, \bibinfo {author} {\bibfnamefont {S.}~\bibnamefont {Sun}}, \bibinfo
  {author} {\bibfnamefont {C.}~\bibnamefont {Guo}}, \bibinfo {author}
  {\bibfnamefont {L.}~\bibnamefont {Wei}}, \bibinfo {author} {\bibfnamefont
  {H.}~\bibnamefont {Tian}}, \bibinfo {author} {\bibfnamefont {H.}~\bibnamefont
  {Yang}}, \bibinfo {author} {\bibfnamefont {G.}~\bibnamefont {Chen}}, \ and\
  \bibinfo {author} {\bibfnamefont {J.}~\bibnamefont {Li}},\ }\href@noop {}
  {\bibfield  {journal} {\bibinfo  {journal} {arxiv}\ } (\bibinfo {year}
  {2016}{\natexlab{b}})},\ \Eprint {http://arxiv.org/abs/1612.07005}
  {1612.07005} \BibitemShut {NoStop}%
\bibitem [{\citenamefont {Chi}\ \emph {et~al.}(2017)\citenamefont {Chi},
  \citenamefont {Li}, \citenamefont {Xie}, \citenamefont {Zhao}, \citenamefont
  {Wang}, \citenamefont {Li}, \citenamefont {Yu}, \citenamefont {Wang},
  \citenamefont {Weng}, \citenamefont {Zhang},\ and\ \citenamefont
  {Wang}}]{Chi2017}%
  \BibitemOpen
  \bibfield  {author} {\bibinfo {author} {\bibfnamefont {S.}~\bibnamefont
  {Chi}}, \bibinfo {author} {\bibfnamefont {Z.}~\bibnamefont {Li}}, \bibinfo
  {author} {\bibfnamefont {Y.}~\bibnamefont {Xie}}, \bibinfo {author}
  {\bibfnamefont {Y.}~\bibnamefont {Zhao}}, \bibinfo {author} {\bibfnamefont
  {Z.}~\bibnamefont {Wang}}, \bibinfo {author} {\bibfnamefont {L.}~\bibnamefont
  {Li}}, \bibinfo {author} {\bibfnamefont {H.}~\bibnamefont {Yu}}, \bibinfo
  {author} {\bibfnamefont {G.}~\bibnamefont {Wang}}, \bibinfo {author}
  {\bibfnamefont {H.}~\bibnamefont {Weng}}, \bibinfo {author} {\bibfnamefont
  {H.}~\bibnamefont {Zhang}}, \ and\ \bibinfo {author} {\bibfnamefont
  {J.}~\bibnamefont {Wang}},\ }\href@noop {} {\bibfield  {journal} {\bibinfo
  {journal} {arxiv}\ } (\bibinfo {year} {2017})},\ \Eprint
  {http://arxiv.org/abs/1705.05086} {1705.05086} \BibitemShut {NoStop}%
\bibitem [{\citenamefont {Moore}\ and\ \citenamefont
  {Orenstein}(2010)}]{Moore2010}%
  \BibitemOpen
  \bibfield  {author} {\bibinfo {author} {\bibfnamefont {J.~E.}\ \bibnamefont
  {Moore}}\ and\ \bibinfo {author} {\bibfnamefont {J.}~\bibnamefont
  {Orenstein}},\ }\href@noop {} {\bibfield  {journal} {\bibinfo  {journal}
  {Phys. Rev. Lett.}\ }\textbf {\bibinfo {volume} {105}},\ \bibinfo {pages}
  {026805} (\bibinfo {year} {2010})}\BibitemShut {NoStop}%
\bibitem [{\citenamefont {Deyo}\ \emph {et~al.}(2009)\citenamefont {Deyo},
  \citenamefont {Golub}, \citenamefont {Ivchenko},\ and\ \citenamefont
  {Spivak}}]{Deyo2009}%
  \BibitemOpen
  \bibfield  {author} {\bibinfo {author} {\bibfnamefont {E.}~\bibnamefont
  {Deyo}}, \bibinfo {author} {\bibfnamefont {L.~E.}\ \bibnamefont {Golub}},
  \bibinfo {author} {\bibfnamefont {E.~L.}\ \bibnamefont {Ivchenko}}, \ and\
  \bibinfo {author} {\bibfnamefont {B.}~\bibnamefont {Spivak}},\ }\href@noop {}
  {\bibfield  {journal} {\bibinfo  {journal} {arxiv}\ } (\bibinfo {year}
  {2009})},\ \Eprint {http://arxiv.org/abs/0904.1917} {0904.1917} \BibitemShut
  {NoStop}%
\bibitem [{\citenamefont {Sipe}\ and\ \citenamefont
  {Shkrebtii}(2000)}]{Sipe2000}%
  \BibitemOpen
  \bibfield  {author} {\bibinfo {author} {\bibfnamefont {J.~E.}\ \bibnamefont
  {Sipe}}\ and\ \bibinfo {author} {\bibfnamefont {A.~I.}\ \bibnamefont
  {Shkrebtii}},\ }\href@noop {} {\bibfield  {journal} {\bibinfo  {journal}
  {Phys. Rev. B}\ }\textbf {\bibinfo {volume} {61}},\ \bibinfo {pages} {5337}
  (\bibinfo {year} {2000})}\BibitemShut {NoStop}%
\bibitem [{\citenamefont {Young}\ and\ \citenamefont
  {Rappe}(2012)}]{Young2012}%
  \BibitemOpen
  \bibfield  {author} {\bibinfo {author} {\bibfnamefont {S.~M.}\ \bibnamefont
  {Young}}\ and\ \bibinfo {author} {\bibfnamefont {A.~M.}\ \bibnamefont
  {Rappe}},\ }\href@noop {} {\bibfield  {journal} {\bibinfo  {journal} {Phys.
  Rev. Lett.}\ }\textbf {\bibinfo {volume} {109}},\ \bibinfo {pages} {116601}
  (\bibinfo {year} {2012})}\BibitemShut {NoStop}%
\bibitem [{\citenamefont {Koepernik}\ and\ \citenamefont
  {Eschrig}(1999)}]{koepernik1999}%
  \BibitemOpen
  \bibfield  {author} {\bibinfo {author} {\bibfnamefont {K.}~\bibnamefont
  {Koepernik}}\ and\ \bibinfo {author} {\bibfnamefont {H.}~\bibnamefont
  {Eschrig}},\ }\href@noop {} {\bibfield  {journal} {\bibinfo  {journal} {Phys.
  Rev. B}\ }\textbf {\bibinfo {volume} {59}},\ \bibinfo {pages} {1743}
  (\bibinfo {year} {1999})}\BibitemShut {NoStop}%
\bibitem [{\citenamefont {Perdew}\ \emph {et~al.}(1996)\citenamefont {Perdew},
  \citenamefont {Burke},\ and\ \citenamefont {Ernzerhof}}]{perdew1996}%
  \BibitemOpen
  \bibfield  {author} {\bibinfo {author} {\bibfnamefont {J.~P.}\ \bibnamefont
  {Perdew}}, \bibinfo {author} {\bibfnamefont {K.}~\bibnamefont {Burke}}, \
  and\ \bibinfo {author} {\bibfnamefont {M.}~\bibnamefont {Ernzerhof}},\
  }\href@noop {} {\bibfield  {journal} {\bibinfo  {journal} {Phys. Rev. Lett.}\
  }\textbf {\bibinfo {volume} {77}},\ \bibinfo {pages} {3865} (\bibinfo {year}
  {1996})}\BibitemShut {NoStop}%
\bibitem [{not()}]{note}%
  \BibitemOpen
  \href@noop {} {}\bibinfo {howpublished} {For the integrals of $D_{xy}$, the
  first Brillouzone was sampled by $k$-grids from $200\times200\times200$ to
  $1000\times1000\times1000$. Satisfactory convergence was achieved for a
  $k$-grid of $800\times800\times800$ for all compounds. Increasing the grid
  size to $1000\times1000\times1000$ only varies the $D_{xy}$ value by no more
  than 5\%.}\BibitemShut {Stop}%
\bibitem [{\citenamefont {Xiao}\ \emph {et~al.}(2010)\citenamefont {Xiao},
  \citenamefont {Chang},\ and\ \citenamefont {Niu}}]{Xiao2010}%
  \BibitemOpen
  \bibfield  {author} {\bibinfo {author} {\bibfnamefont {D.}~\bibnamefont
  {Xiao}}, \bibinfo {author} {\bibfnamefont {M.-C.}\ \bibnamefont {Chang}}, \
  and\ \bibinfo {author} {\bibfnamefont {Q.}~\bibnamefont {Niu}},\ }\href@noop
  {} {\bibfield  {journal} {\bibinfo  {journal} {Rev. Mod. Phys.}\ }\textbf
  {\bibinfo {volume} {82}},\ \bibinfo {pages} {1959} (\bibinfo {year}
  {2010})}\BibitemShut {NoStop}%
\bibitem [{\citenamefont {Arnold}\ \emph
  {et~al.}(2016{\natexlab{a}})\citenamefont {Arnold}, \citenamefont {Naumann},
  \citenamefont {Wu}, \citenamefont {Sun}, \citenamefont {Schmidt},
  \citenamefont {Borrmann}, \citenamefont {Felser}, \citenamefont {Yan},\ and\
  \citenamefont {Hassinger}}]{Arnold2016FS}%
  \BibitemOpen
  \bibfield  {author} {\bibinfo {author} {\bibfnamefont {F.}~\bibnamefont
  {Arnold}}, \bibinfo {author} {\bibfnamefont {M.}~\bibnamefont {Naumann}},
  \bibinfo {author} {\bibfnamefont {S.~C.}\ \bibnamefont {Wu}}, \bibinfo
  {author} {\bibfnamefont {Y.}~\bibnamefont {Sun}}, \bibinfo {author}
  {\bibfnamefont {M.}~\bibnamefont {Schmidt}}, \bibinfo {author} {\bibfnamefont
  {H.}~\bibnamefont {Borrmann}}, \bibinfo {author} {\bibfnamefont
  {C.}~\bibnamefont {Felser}}, \bibinfo {author} {\bibfnamefont
  {B.}~\bibnamefont {Yan}}, \ and\ \bibinfo {author} {\bibfnamefont
  {E.}~\bibnamefont {Hassinger}},\ }\href@noop {} {\bibfield  {journal}
  {\bibinfo  {journal} {Phys. Rev. Lett.}\ }\textbf {\bibinfo {volume} {117}},\
  \bibinfo {pages} {146401} (\bibinfo {year} {2016}{\natexlab{a}})}\BibitemShut
  {NoStop}%
\bibitem [{\citenamefont {Wang}\ \emph {et~al.}(2016)\citenamefont {Wang},
  \citenamefont {Gresch}, \citenamefont {Soluyanov}, \citenamefont {Xie},
  \citenamefont {Kushwaha}, \citenamefont {Dai}, \citenamefont {Troyer},
  \citenamefont {Cava},\ and\ \citenamefont {Bernevig}}]{Wang2016MoTe2}%
  \BibitemOpen
  \bibfield  {author} {\bibinfo {author} {\bibfnamefont {Z.}~\bibnamefont
  {Wang}}, \bibinfo {author} {\bibfnamefont {D.}~\bibnamefont {Gresch}},
  \bibinfo {author} {\bibfnamefont {A.~A.}\ \bibnamefont {Soluyanov}}, \bibinfo
  {author} {\bibfnamefont {W.}~\bibnamefont {Xie}}, \bibinfo {author}
  {\bibfnamefont {S.}~\bibnamefont {Kushwaha}}, \bibinfo {author}
  {\bibfnamefont {X.}~\bibnamefont {Dai}}, \bibinfo {author} {\bibfnamefont
  {M.}~\bibnamefont {Troyer}}, \bibinfo {author} {\bibfnamefont {R.~J.}\
  \bibnamefont {Cava}}, \ and\ \bibinfo {author} {\bibfnamefont {B.~A.}\
  \bibnamefont {Bernevig}},\ }\href@noop {} {\bibfield  {journal} {\bibinfo
  {journal} {Phys. Rev. Lett.}\ }\textbf {\bibinfo {volume} {117}},\ \bibinfo
  {pages} {056805} (\bibinfo {year} {2016})}\BibitemShut {NoStop}%
\bibitem [{\citenamefont {Bruno}\ \emph {et~al.}(2016)\citenamefont {Bruno},
  \citenamefont {Tamai}, \citenamefont {Wu}, \citenamefont {Cucchi},
  \citenamefont {Barreteau}, \citenamefont {de~la Torre}, \citenamefont
  {Walker}, \citenamefont {Ricc{\`o}}, \citenamefont {Wang}, \citenamefont
  {Kim}, \citenamefont {Hoesch}, \citenamefont {Shi}, \citenamefont {Plumb},
  \citenamefont {Giannini}, \citenamefont {Soluyanov},\ and\ \citenamefont
  {Baumberger}}]{Bruno2016}%
  \BibitemOpen
  \bibfield  {author} {\bibinfo {author} {\bibfnamefont {F.~Y.}\ \bibnamefont
  {Bruno}}, \bibinfo {author} {\bibfnamefont {A.}~\bibnamefont {Tamai}},
  \bibinfo {author} {\bibfnamefont {Q.~S.}\ \bibnamefont {Wu}}, \bibinfo
  {author} {\bibfnamefont {I.}~\bibnamefont {Cucchi}}, \bibinfo {author}
  {\bibfnamefont {C.}~\bibnamefont {Barreteau}}, \bibinfo {author}
  {\bibfnamefont {A.}~\bibnamefont {de~la Torre}}, \bibinfo {author}
  {\bibfnamefont {S.~M.}\ \bibnamefont {Walker}}, \bibinfo {author}
  {\bibfnamefont {S.}~\bibnamefont {Ricc{\`o}}}, \bibinfo {author}
  {\bibfnamefont {Z.}~\bibnamefont {Wang}}, \bibinfo {author} {\bibfnamefont
  {T.~K.}\ \bibnamefont {Kim}}, \bibinfo {author} {\bibfnamefont
  {M.}~\bibnamefont {Hoesch}}, \bibinfo {author} {\bibfnamefont
  {M.}~\bibnamefont {Shi}}, \bibinfo {author} {\bibfnamefont {N.~C.}\
  \bibnamefont {Plumb}}, \bibinfo {author} {\bibfnamefont {E.}~\bibnamefont
  {Giannini}}, \bibinfo {author} {\bibfnamefont {A.~A.}\ \bibnamefont
  {Soluyanov}}, \ and\ \bibinfo {author} {\bibfnamefont {F.}~\bibnamefont
  {Baumberger}},\ }\href@noop {} {\bibfield  {journal} {\bibinfo  {journal}
  {Phys. Rev. B}\ }\textbf {\bibinfo {volume} {94}},\ \bibinfo {pages} {121112}
  (\bibinfo {year} {2016})}\BibitemShut {NoStop}%
\bibitem [{\citenamefont {Wu}\ \emph {et~al.}(2017{\natexlab{b}})\citenamefont
  {Wu}, \citenamefont {Sun}, \citenamefont {Claudia},\ and\ \citenamefont
  {Yan}}]{Wu2017}%
  \BibitemOpen
  \bibfield  {author} {\bibinfo {author} {\bibfnamefont {S.-C.}\ \bibnamefont
  {Wu}}, \bibinfo {author} {\bibfnamefont {Y.}~\bibnamefont {Sun}}, \bibinfo
  {author} {\bibfnamefont {F.}~\bibnamefont {Claudia}}, \ and\ \bibinfo
  {author} {\bibfnamefont {B.}~\bibnamefont {Yan}},\ }\href@noop {} {\bibfield
  {journal} {\bibinfo  {journal} {arXiv}\ } (\bibinfo {year}
  {2017}{\natexlab{b}})},\ \Eprint {http://arxiv.org/abs/1708.07002}
  {1708.07002} \BibitemShut {NoStop}%
\bibitem [{\citenamefont {Klotz}\ \emph {et~al.}(2016)\citenamefont {Klotz},
  \citenamefont {Wu}, \citenamefont {Shekhar}, \citenamefont {Sun},
  \citenamefont {Schmidt}, \citenamefont {Nicklas}, \citenamefont {Baenitz},
  \citenamefont {Uhlarz}, \citenamefont {Wosnitza}, \citenamefont {Felser},\
  and\ \citenamefont {Yan}}]{Klotz2016}%
  \BibitemOpen
  \bibfield  {author} {\bibinfo {author} {\bibfnamefont {J.}~\bibnamefont
  {Klotz}}, \bibinfo {author} {\bibfnamefont {S.-C.}\ \bibnamefont {Wu}},
  \bibinfo {author} {\bibfnamefont {C.}~\bibnamefont {Shekhar}}, \bibinfo
  {author} {\bibfnamefont {Y.}~\bibnamefont {Sun}}, \bibinfo {author}
  {\bibfnamefont {M.}~\bibnamefont {Schmidt}}, \bibinfo {author} {\bibfnamefont
  {M.}~\bibnamefont {Nicklas}}, \bibinfo {author} {\bibfnamefont
  {M.}~\bibnamefont {Baenitz}}, \bibinfo {author} {\bibfnamefont
  {M.}~\bibnamefont {Uhlarz}}, \bibinfo {author} {\bibfnamefont
  {J.}~\bibnamefont {Wosnitza}}, \bibinfo {author} {\bibfnamefont
  {C.}~\bibnamefont {Felser}}, \ and\ \bibinfo {author} {\bibfnamefont
  {B.}~\bibnamefont {Yan}},\ }\href@noop {} {\bibfield  {journal} {\bibinfo
  {journal} {Phys. Rev. B}\ }\textbf {\bibinfo {volume} {93}},\ \bibinfo
  {pages} {121105} (\bibinfo {year} {2016})}\BibitemShut {NoStop}%
\bibitem [{\citenamefont {dos Reis}\ \emph {et~al.}(2016)\citenamefont {dos
  Reis}, \citenamefont {Wu}, \citenamefont {Sun}, \citenamefont {Ajeesh},
  \citenamefont {Shekhar}, \citenamefont {Schmidt}, \citenamefont {Felser},
  \citenamefont {Yan},\ and\ \citenamefont {Nicklas}}]{Reis2016}%
  \BibitemOpen
  \bibfield  {author} {\bibinfo {author} {\bibfnamefont {R.~D.}\ \bibnamefont
  {dos Reis}}, \bibinfo {author} {\bibfnamefont {S.~C.}\ \bibnamefont {Wu}},
  \bibinfo {author} {\bibfnamefont {Y.}~\bibnamefont {Sun}}, \bibinfo {author}
  {\bibfnamefont {M.~O.}\ \bibnamefont {Ajeesh}}, \bibinfo {author}
  {\bibfnamefont {C.}~\bibnamefont {Shekhar}}, \bibinfo {author} {\bibfnamefont
  {M.}~\bibnamefont {Schmidt}}, \bibinfo {author} {\bibfnamefont
  {C.}~\bibnamefont {Felser}}, \bibinfo {author} {\bibfnamefont
  {B.}~\bibnamefont {Yan}}, \ and\ \bibinfo {author} {\bibfnamefont
  {M.}~\bibnamefont {Nicklas}},\ }\href@noop {} {\bibfield  {journal} {\bibinfo
   {journal} {Phys. Rev. B}\ }\textbf {\bibinfo {volume} {93}},\ \bibinfo
  {pages} {205102} (\bibinfo {year} {2016})}\BibitemShut {NoStop}%
\bibitem [{\citenamefont {Belopolski}\ \emph {et~al.}(2016)\citenamefont
  {Belopolski}, \citenamefont {Sanchez}, \citenamefont {Ishida}, \citenamefont
  {Pan}, \citenamefont {Yu}, \citenamefont {Xu}, \citenamefont {Chang},
  \citenamefont {Chang}, \citenamefont {Zheng}, \citenamefont {Alidoust},
  \citenamefont {Bian}, \citenamefont {Neupane}, \citenamefont {Huang},
  \citenamefont {Lee}, \citenamefont {Song}, \citenamefont {Bu}, \citenamefont
  {Wang}, \citenamefont {Li}, \citenamefont {Eda}, \citenamefont {Jeng},
  \citenamefont {Kondo}, \citenamefont {Lin}, \citenamefont {Liu},
  \citenamefont {Song}, \citenamefont {Shin},\ and\ \citenamefont
  {Hasan}}]{Belopolski2016}%
  \BibitemOpen
  \bibfield  {author} {\bibinfo {author} {\bibfnamefont {I.}~\bibnamefont
  {Belopolski}}, \bibinfo {author} {\bibfnamefont {D.~S.}\ \bibnamefont
  {Sanchez}}, \bibinfo {author} {\bibfnamefont {Y.}~\bibnamefont {Ishida}},
  \bibinfo {author} {\bibfnamefont {X.}~\bibnamefont {Pan}}, \bibinfo {author}
  {\bibfnamefont {P.}~\bibnamefont {Yu}}, \bibinfo {author} {\bibfnamefont
  {S.-Y.}\ \bibnamefont {Xu}}, \bibinfo {author} {\bibfnamefont
  {G.}~\bibnamefont {Chang}}, \bibinfo {author} {\bibfnamefont {T.-R.}\
  \bibnamefont {Chang}}, \bibinfo {author} {\bibfnamefont {H.}~\bibnamefont
  {Zheng}}, \bibinfo {author} {\bibfnamefont {N.}~\bibnamefont {Alidoust}},
  \bibinfo {author} {\bibfnamefont {G.}~\bibnamefont {Bian}}, \bibinfo {author}
  {\bibfnamefont {M.}~\bibnamefont {Neupane}}, \bibinfo {author} {\bibfnamefont
  {S.-M.}\ \bibnamefont {Huang}}, \bibinfo {author} {\bibfnamefont {C.-C.}\
  \bibnamefont {Lee}}, \bibinfo {author} {\bibfnamefont {Y.}~\bibnamefont
  {Song}}, \bibinfo {author} {\bibfnamefont {H.}~\bibnamefont {Bu}}, \bibinfo
  {author} {\bibfnamefont {G.}~\bibnamefont {Wang}}, \bibinfo {author}
  {\bibfnamefont {S.}~\bibnamefont {Li}}, \bibinfo {author} {\bibfnamefont
  {G.}~\bibnamefont {Eda}}, \bibinfo {author} {\bibfnamefont {H.-T.}\
  \bibnamefont {Jeng}}, \bibinfo {author} {\bibfnamefont {T.}~\bibnamefont
  {Kondo}}, \bibinfo {author} {\bibfnamefont {H.}~\bibnamefont {Lin}}, \bibinfo
  {author} {\bibfnamefont {Z.}~\bibnamefont {Liu}}, \bibinfo {author}
  {\bibfnamefont {F.}~\bibnamefont {Song}}, \bibinfo {author} {\bibfnamefont
  {S.}~\bibnamefont {Shin}}, \ and\ \bibinfo {author} {\bibfnamefont {M.~Z.}\
  \bibnamefont {Hasan}},\ }\href@noop {} {\bibfield  {journal} {\bibinfo
  {journal} {Nat. Commun.}\ }\textbf {\bibinfo {volume} {7}},\ \bibinfo {pages}
  {ncomms13643} (\bibinfo {year} {2016})}\BibitemShut {NoStop}%
\bibitem [{\citenamefont {Einaga}\ \emph {et~al.}(2017)\citenamefont {Einaga},
  \citenamefont {Shimizu}, \citenamefont {Hu}, \citenamefont {Mao},\ and\
  \citenamefont {Politano}}]{Einaga2017}%
  \BibitemOpen
  \bibfield  {author} {\bibinfo {author} {\bibfnamefont {M.}~\bibnamefont
  {Einaga}}, \bibinfo {author} {\bibfnamefont {K.}~\bibnamefont {Shimizu}},
  \bibinfo {author} {\bibfnamefont {J.}~\bibnamefont {Hu}}, \bibinfo {author}
  {\bibfnamefont {Z.~Q.}\ \bibnamefont {Mao}}, \ and\ \bibinfo {author}
  {\bibfnamefont {A.}~\bibnamefont {Politano}},\ }\href@noop {} {\bibfield
  {journal} {\bibinfo  {journal} {Phys. Status Solidi RRL}\ }\textbf {\bibinfo
  {volume} {11}},\ \bibinfo {pages} {1700182} (\bibinfo {year}
  {2017})}\BibitemShut {NoStop}%
\bibitem [{\citenamefont {Nagaosa}\ \emph {et~al.}(2010)\citenamefont
  {Nagaosa}, \citenamefont {Sinova}, \citenamefont {Onoda}, \citenamefont
  {MacDonald},\ and\ \citenamefont {Ong}}]{Nagaosa2010}%
  \BibitemOpen
  \bibfield  {author} {\bibinfo {author} {\bibfnamefont {N.}~\bibnamefont
  {Nagaosa}}, \bibinfo {author} {\bibfnamefont {J.}~\bibnamefont {Sinova}},
  \bibinfo {author} {\bibfnamefont {S.}~\bibnamefont {Onoda}}, \bibinfo
  {author} {\bibfnamefont {A.~H.}\ \bibnamefont {MacDonald}}, \ and\ \bibinfo
  {author} {\bibfnamefont {N.~P.}\ \bibnamefont {Ong}},\ }\href@noop {}
  {\bibfield  {journal} {\bibinfo  {journal} {Reviews of Modern Physics}\
  }\textbf {\bibinfo {volume} {82}},\ \bibinfo {pages} {1539} (\bibinfo {year}
  {2010})}\BibitemShut {NoStop}%
\bibitem [{\citenamefont {Shekhar}\ \emph {et~al.}(2015)\citenamefont
  {Shekhar}, \citenamefont {Nayak}, \citenamefont {Sun}, \citenamefont
  {Schmidt}, \citenamefont {Nicklas}, \citenamefont {Leermakers}, \citenamefont
  {Zeitler}, \citenamefont {Skourski}, \citenamefont {Wosnitza}, \citenamefont
  {Liu}, \citenamefont {Chen}, \citenamefont {Schnelle}, \citenamefont
  {Borrmann}, \citenamefont {Grin}, \citenamefont {Felser},\ and\ \citenamefont
  {Yan}}]{Shekhar2015}%
  \BibitemOpen
  \bibfield  {author} {\bibinfo {author} {\bibfnamefont {C.}~\bibnamefont
  {Shekhar}}, \bibinfo {author} {\bibfnamefont {A.~K.}\ \bibnamefont {Nayak}},
  \bibinfo {author} {\bibfnamefont {Y.}~\bibnamefont {Sun}}, \bibinfo {author}
  {\bibfnamefont {M.}~\bibnamefont {Schmidt}}, \bibinfo {author} {\bibfnamefont
  {M.}~\bibnamefont {Nicklas}}, \bibinfo {author} {\bibfnamefont
  {I.}~\bibnamefont {Leermakers}}, \bibinfo {author} {\bibfnamefont
  {U.}~\bibnamefont {Zeitler}}, \bibinfo {author} {\bibfnamefont
  {Y.}~\bibnamefont {Skourski}}, \bibinfo {author} {\bibfnamefont
  {J.}~\bibnamefont {Wosnitza}}, \bibinfo {author} {\bibfnamefont
  {Z.}~\bibnamefont {Liu}}, \bibinfo {author} {\bibfnamefont {Y.}~\bibnamefont
  {Chen}}, \bibinfo {author} {\bibfnamefont {W.}~\bibnamefont {Schnelle}},
  \bibinfo {author} {\bibfnamefont {H.}~\bibnamefont {Borrmann}}, \bibinfo
  {author} {\bibfnamefont {Y.}~\bibnamefont {Grin}}, \bibinfo {author}
  {\bibfnamefont {C.}~\bibnamefont {Felser}}, \ and\ \bibinfo {author}
  {\bibfnamefont {B.}~\bibnamefont {Yan}},\ }\href@noop {} {\bibfield
  {journal} {\bibinfo  {journal} {Nat. Phys.}\ }\textbf {\bibinfo {volume}
  {11}},\ \bibinfo {pages} {645} (\bibinfo {year} {2015})}\BibitemShut
  {NoStop}%
\bibitem [{\citenamefont {Arnold}\ \emph
  {et~al.}(2016{\natexlab{b}})\citenamefont {Arnold}, \citenamefont {Shekhar},
  \citenamefont {Wu}, \citenamefont {Sun}, \citenamefont {dos Reis},
  \citenamefont {Kumar}, \citenamefont {Naumann}, \citenamefont {Ajeesh},
  \citenamefont {Schmidt}, \citenamefont {Grushin}, \citenamefont {Bardarson},
  \citenamefont {Baenitz}, \citenamefont {Sokolov}, \citenamefont {Borrmann},
  \citenamefont {Nicklas}, \citenamefont {Felser}, \citenamefont {Hassinger},\
  and\ \citenamefont {Yan}}]{Arnold2016}%
  \BibitemOpen
  \bibfield  {author} {\bibinfo {author} {\bibfnamefont {F.}~\bibnamefont
  {Arnold}}, \bibinfo {author} {\bibfnamefont {C.}~\bibnamefont {Shekhar}},
  \bibinfo {author} {\bibfnamefont {S.-C.}\ \bibnamefont {Wu}}, \bibinfo
  {author} {\bibfnamefont {Y.}~\bibnamefont {Sun}}, \bibinfo {author}
  {\bibfnamefont {R.~D.}\ \bibnamefont {dos Reis}}, \bibinfo {author}
  {\bibfnamefont {N.}~\bibnamefont {Kumar}}, \bibinfo {author} {\bibfnamefont
  {M.}~\bibnamefont {Naumann}}, \bibinfo {author} {\bibfnamefont {M.~O.}\
  \bibnamefont {Ajeesh}}, \bibinfo {author} {\bibfnamefont {M.}~\bibnamefont
  {Schmidt}}, \bibinfo {author} {\bibfnamefont {A.~G.}\ \bibnamefont
  {Grushin}}, \bibinfo {author} {\bibfnamefont {J.~H.}\ \bibnamefont
  {Bardarson}}, \bibinfo {author} {\bibfnamefont {M.}~\bibnamefont {Baenitz}},
  \bibinfo {author} {\bibfnamefont {D.}~\bibnamefont {Sokolov}}, \bibinfo
  {author} {\bibfnamefont {H.}~\bibnamefont {Borrmann}}, \bibinfo {author}
  {\bibfnamefont {M.}~\bibnamefont {Nicklas}}, \bibinfo {author} {\bibfnamefont
  {C.}~\bibnamefont {Felser}}, \bibinfo {author} {\bibfnamefont
  {E.}~\bibnamefont {Hassinger}}, \ and\ \bibinfo {author} {\bibfnamefont
  {B.}~\bibnamefont {Yan}},\ }\href@noop {} {\bibfield  {journal} {\bibinfo
  {journal} {Nat. Commun.}\ }\textbf {\bibinfo {volume} {7}},\ \bibinfo {pages}
  {11615} (\bibinfo {year} {2016}{\natexlab{b}})}\BibitemShut {NoStop}%
\bibitem [{\citenamefont {Zhang}\ \emph {et~al.}(2016)\citenamefont {Zhang},
  \citenamefont {Xu}, \citenamefont {Belopolski}, \citenamefont {Yuan},
  \citenamefont {Lin}, \citenamefont {Tong}, \citenamefont {Bian},
  \citenamefont {Alidoust}, \citenamefont {Lee}, \citenamefont {Huang},
  \citenamefont {Chang}, \citenamefont {Chang}, \citenamefont {Hsu},
  \citenamefont {Jeng}, \citenamefont {Neupane}, \citenamefont {Sanchez},
  \citenamefont {Zheng}, \citenamefont {Wang}, \citenamefont {Lin},
  \citenamefont {Zhang}, \citenamefont {Lu}, \citenamefont {Shen},
  \citenamefont {Neupert}, \citenamefont {Zahid~Hasan},\ and\ \citenamefont
  {Jia}}]{Zhang2016ABJ}%
  \BibitemOpen
  \bibfield  {author} {\bibinfo {author} {\bibfnamefont {C.-L.}\ \bibnamefont
  {Zhang}}, \bibinfo {author} {\bibfnamefont {S.-Y.}\ \bibnamefont {Xu}},
  \bibinfo {author} {\bibfnamefont {I.}~\bibnamefont {Belopolski}}, \bibinfo
  {author} {\bibfnamefont {Z.}~\bibnamefont {Yuan}}, \bibinfo {author}
  {\bibfnamefont {Z.}~\bibnamefont {Lin}}, \bibinfo {author} {\bibfnamefont
  {B.}~\bibnamefont {Tong}}, \bibinfo {author} {\bibfnamefont {G.}~\bibnamefont
  {Bian}}, \bibinfo {author} {\bibfnamefont {N.}~\bibnamefont {Alidoust}},
  \bibinfo {author} {\bibfnamefont {C.-C.}\ \bibnamefont {Lee}}, \bibinfo
  {author} {\bibfnamefont {S.-M.}\ \bibnamefont {Huang}}, \bibinfo {author}
  {\bibfnamefont {T.-R.}\ \bibnamefont {Chang}}, \bibinfo {author}
  {\bibfnamefont {G.}~\bibnamefont {Chang}}, \bibinfo {author} {\bibfnamefont
  {C.-H.}\ \bibnamefont {Hsu}}, \bibinfo {author} {\bibfnamefont {H.-T.}\
  \bibnamefont {Jeng}}, \bibinfo {author} {\bibfnamefont {M.}~\bibnamefont
  {Neupane}}, \bibinfo {author} {\bibfnamefont {D.~S.}\ \bibnamefont
  {Sanchez}}, \bibinfo {author} {\bibfnamefont {H.}~\bibnamefont {Zheng}},
  \bibinfo {author} {\bibfnamefont {J.}~\bibnamefont {Wang}}, \bibinfo {author}
  {\bibfnamefont {H.}~\bibnamefont {Lin}}, \bibinfo {author} {\bibfnamefont
  {C.}~\bibnamefont {Zhang}}, \bibinfo {author} {\bibfnamefont {H.-Z.}\
  \bibnamefont {Lu}}, \bibinfo {author} {\bibfnamefont {S.-Q.}\ \bibnamefont
  {Shen}}, \bibinfo {author} {\bibfnamefont {T.}~\bibnamefont {Neupert}},
  \bibinfo {author} {\bibfnamefont {M.}~\bibnamefont {Zahid~Hasan}}, \ and\
  \bibinfo {author} {\bibfnamefont {S.}~\bibnamefont {Jia}},\ }\href@noop {}
  {\bibfield  {journal} {\bibinfo  {journal} {Nat. Commun.}\ }\textbf {\bibinfo
  {volume} {7}},\ \bibinfo {pages} {10735} (\bibinfo {year}
  {2016})}\BibitemShut {NoStop}%
\bibitem [{\citenamefont {Huang}\ \emph
  {et~al.}(2015{\natexlab{b}})\citenamefont {Huang}, \citenamefont {Zhao},
  \citenamefont {Long}, \citenamefont {Wang}, \citenamefont {Chen},
  \citenamefont {Yang}, \citenamefont {Liang}, \citenamefont {Xue},
  \citenamefont {Weng}, \citenamefont {Fang}, \citenamefont {Dai},\ and\
  \citenamefont {Chen}}]{Huang2015anomaly}%
  \BibitemOpen
  \bibfield  {author} {\bibinfo {author} {\bibfnamefont {X.}~\bibnamefont
  {Huang}}, \bibinfo {author} {\bibfnamefont {L.}~\bibnamefont {Zhao}},
  \bibinfo {author} {\bibfnamefont {Y.}~\bibnamefont {Long}}, \bibinfo {author}
  {\bibfnamefont {P.}~\bibnamefont {Wang}}, \bibinfo {author} {\bibfnamefont
  {D.}~\bibnamefont {Chen}}, \bibinfo {author} {\bibfnamefont {Z.}~\bibnamefont
  {Yang}}, \bibinfo {author} {\bibfnamefont {H.}~\bibnamefont {Liang}},
  \bibinfo {author} {\bibfnamefont {M.}~\bibnamefont {Xue}}, \bibinfo {author}
  {\bibfnamefont {H.}~\bibnamefont {Weng}}, \bibinfo {author} {\bibfnamefont
  {Z.}~\bibnamefont {Fang}}, \bibinfo {author} {\bibfnamefont {X.}~\bibnamefont
  {Dai}}, \ and\ \bibinfo {author} {\bibfnamefont {G.}~\bibnamefont {Chen}},\
  }\href@noop {} {\bibfield  {journal} {\bibinfo  {journal} {Phys. Rev. X}\
  }\textbf {\bibinfo {volume} {5}},\ \bibinfo {pages} {031023} (\bibinfo {year}
  {2015}{\natexlab{b}})}\BibitemShut {NoStop}%
\bibitem [{\citenamefont {Qi}\ \emph {et~al.}(2016)\citenamefont {Qi},
  \citenamefont {Naumov}, \citenamefont {Ali}, \citenamefont {Rajamathi},
  \citenamefont {Schnelle}, \citenamefont {Barkalov}, \citenamefont {Hanfland},
  \citenamefont {Wu}, \citenamefont {Shekhar}, \citenamefont {Sun},
  \citenamefont {Su{\ss}}, \citenamefont {Schmidt}, \citenamefont {Schwarz},
  \citenamefont {Pippel}, \citenamefont {Werner}, \citenamefont {Hillebrand},
  \citenamefont {Forster}, \citenamefont {Kampert}, \citenamefont {Parkin},
  \citenamefont {Cava}, \citenamefont {Felser}, \citenamefont {Yan},\ and\
  \citenamefont {Medvedev}}]{Qi2016MoTe2}%
  \BibitemOpen
  \bibfield  {author} {\bibinfo {author} {\bibfnamefont {Y.}~\bibnamefont
  {Qi}}, \bibinfo {author} {\bibfnamefont {P.~G.}\ \bibnamefont {Naumov}},
  \bibinfo {author} {\bibfnamefont {M.~N.}\ \bibnamefont {Ali}}, \bibinfo
  {author} {\bibfnamefont {C.~R.}\ \bibnamefont {Rajamathi}}, \bibinfo {author}
  {\bibfnamefont {W.}~\bibnamefont {Schnelle}}, \bibinfo {author}
  {\bibfnamefont {O.}~\bibnamefont {Barkalov}}, \bibinfo {author}
  {\bibfnamefont {M.}~\bibnamefont {Hanfland}}, \bibinfo {author}
  {\bibfnamefont {S.-C.}\ \bibnamefont {Wu}}, \bibinfo {author} {\bibfnamefont
  {C.}~\bibnamefont {Shekhar}}, \bibinfo {author} {\bibfnamefont
  {Y.}~\bibnamefont {Sun}}, \bibinfo {author} {\bibfnamefont {V.}~\bibnamefont
  {Su{\ss}}}, \bibinfo {author} {\bibfnamefont {M.}~\bibnamefont {Schmidt}},
  \bibinfo {author} {\bibfnamefont {U.}~\bibnamefont {Schwarz}}, \bibinfo
  {author} {\bibfnamefont {E.}~\bibnamefont {Pippel}}, \bibinfo {author}
  {\bibfnamefont {P.}~\bibnamefont {Werner}}, \bibinfo {author} {\bibfnamefont
  {R.}~\bibnamefont {Hillebrand}}, \bibinfo {author} {\bibfnamefont
  {T.}~\bibnamefont {Forster}}, \bibinfo {author} {\bibfnamefont
  {E.}~\bibnamefont {Kampert}}, \bibinfo {author} {\bibfnamefont
  {S.}~\bibnamefont {Parkin}}, \bibinfo {author} {\bibfnamefont {R.~J.}\
  \bibnamefont {Cava}}, \bibinfo {author} {\bibfnamefont {C.}~\bibnamefont
  {Felser}}, \bibinfo {author} {\bibfnamefont {B.}~\bibnamefont {Yan}}, \ and\
  \bibinfo {author} {\bibfnamefont {S.~A.}\ \bibnamefont {Medvedev}},\
  }\href@noop {} {\bibfield  {journal} {\bibinfo  {journal} {Nat. Commun.}\
  }\textbf {\bibinfo {volume} {7}},\ \bibinfo {pages} {11038} (\bibinfo {year}
  {2016})}\BibitemShut {NoStop}%
\end{thebibliography}
%

\end{document}